\documentclass[hidelinks]{article}
\usepackage{caption}
\usepackage{amsmath} 
\usepackage{amssymb}
\usepackage{graphicx}
\usepackage{natbib}
\usepackage{indentfirst}
\usepackage{xcolor, soul}
\usepackage[hidelinks]{hyperref}
\usepackage{authblk}
\usepackage[english]{babel}

\newcommand{\bx}{\mathbf{x}}

\newcommand{\bF}{\mathbf{F}}

\newcommand{\bD}{\mathbf{D}}

\newcommand{\bI}{\mathbf{I}}

\newcommand{\br}{\mathbf{r}}

\newcommand{\Fig}[1]{Fig.~\ref{#1}}

\newcommand{\Eq}[1]{Eq.~(\ref{#1})}
\newcommand{\Eqs}[2]{Eqs.~(\ref{#1}) and (\ref{#2})}

\newcommand{\fcvx}{{f_{\text{cvx}}}}

\newcommand{\fcvxI}{{f^I_{\text{cvx}}}}
\newcommand{\fcvxE}{{f^E_{\text{cvx}}}}

\title{Approximating nonlinear functions with latent boundaries in low-rank excitatory-inhibitory spiking networks}

\author{William F. Podlaski \\ \small william.podlaski@research.fchampalimaud.org 
\and
Christian K. Machens  \\ \small christian.machens@neuro.fchampalimaud.org
}

\affil{Champalimaud Neuroscience Programme, Champalimaud Foundation,\\ Avenida Brasilia, 1400-038 Lisbon, Portugal}

\begin{document}

\maketitle

% \newpage

\section*{Abstract}
Deep feedforward and recurrent rate-based neural networks have become successful functional models of the brain, but they neglect obvious biological details such as spikes and Dale's law. Here we argue that these details are crucial in order to understand how real neural circuits operate. Towards this aim, we put forth a new framework for spike-based computation in low-rank excitatory-inhibitory spiking networks. By considering populations with rank-1 connectivity, we cast each neuron's spiking threshold as a boundary in a low-dimensional input-output space. We then show how the combined thresholds of a population of inhibitory neurons form a stable boundary in this space, and those of a population of excitatory neurons form an unstable boundary. Combining the two boundaries results in a rank-2 excitatory-inhibitory (EI) network with inhibition-stabilized dynamics at the intersection of the two boundaries. The computation of the resulting networks can be understood as the difference of two convex functions and is thereby capable of approximating arbitrary non-linear input-output mappings. We demonstrate several properties of these networks, including noise suppression and amplification, irregular activity and synaptic balance, as well as how they relate to rate network dynamics in the limit that the boundary becomes soft. Finally, while our work focuses on small networks (5-50 neurons), we discuss potential avenues for scaling up to much larger networks. Overall, our work proposes a new perspective on spiking networks that may serve as a starting point for a mechanistic understanding of biological spike-based computation.

\newpage

\tableofcontents

\newpage

%%%%%%%%%%%%%%%%%%%%%%%%%%%%%%%%%%%%%%%%%%%%%%%%%%%%%
%%%%%%%%%%%%%%%%%%%%%%%%%%%%%%%%%%%%%%%%%%%%%%%%%%%%%
\section{Introduction}

The neural circuits of the brain are unbelievably complex. Yet when it comes to studying how they compute, we often resort to highly simplified network models composed of neurons with graded activation functions, i.e., rate neurons. The resulting rate networks have become the standard models of feedforward sensory processing \citep{yamins2016using, lindsay2021convolutional} and recurrent task dynamics \citep{sussillo2014neural, barak2017recurrent} and capture many aspects of neural circuits surprisingly well.

Of course, a mechanistic understanding of biological computation must eventually bridge back to the details of real circuits \citep{bernaez2022incorporate}. However, this task has proven surprisingly hard. Indeed, the more biologically detailed a network model is, the more difficult it tends to be to constrain and interpret \citep{eliasmith2014use}. We argue here that a fundamental part of this problem lies in two computational concepts of rate networks that are mismatched with biology.

\begin{figure}[t!]
\centering
  \includegraphics[scale=1]{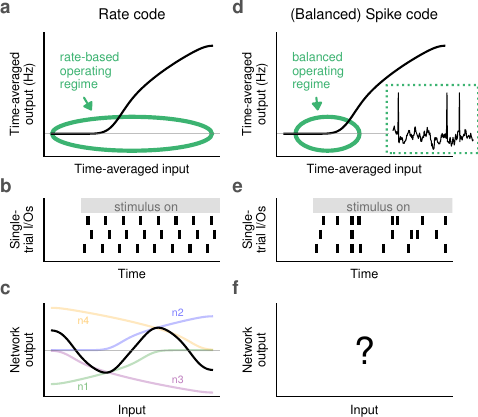}
  \caption{\label{puzzle-fig}
    The operating regimes and sign constraints of rate (\textbf{a-c}) and balanced spike (\textbf{d-f}) codes are fundamentally different. \textbf{a}: A rate-coding perspective utilizes the time-averaged frequency-input (f-I) curve (shown in black for a leaky integrate-and-fire (LIF) neuron receiving noisy input) as a basis; such neurons may be activated far above or below threshold (green ellipse). \textbf{b}: In the absence of additional noise, rate-coding neurons fire regularly in response to an input stimulus (3 trials shown) with a rate according to the f-I curve. \textbf{c}: Function approximation through linear-nonlinear mappings may be achieved in rate-based spiking networks (e.g., \cite{eliasmith2003neural}); here, the black output curve is composed of the weighted sum of the four spiking neuron f-I curves (n1, green; n2, blue; n3, purple; n4, orange), with each neuron having a positive or negative output weight. \textbf{d}: A balanced-input perspective of a spiking neuron suggests a more localized operating regime with mean input below or at the neuron's threshold (green ellipse), and thus does not follow the f-I curve (black); this balance is due to the equal strength of excitatory and inhibitory inputs. \textbf{e}: This regime explains the irregular firing and trial-to-trial variability seen in cortex. \textbf{f}: It is less well-understood how such a balanced, Daleian regime may explain biological computation.}
\end{figure}

The first concept is that of the `feature detector' \citep{martin1994brief}, and pertains to the difference between rate-based and spike-based coding. In a nutshell, rate neurons operate in a regime of depolarized inputs, i.e., they are activated far beyond threshold whenever the neuron's input pattern matches its pattern of synaptic weights. In the spiking domain, unless a large amount of external noise is added, a direct translation of this idea leads to regular spike trains (\cite{eliasmith2003neural}; \Fig{puzzle-fig}a-c). While this may be an accurate description of some biological neurons (e.g., those at the sensory periphery), cortical neurons often fire spikes irregularly \citep{softky1993highly}. The cause of this irregularity is that such neurons operate in a fluctuation-driven regime, in which excitatory and inhibitory input currents balance each other on average \citep{van1996chaos, shadlen1998variable, haider2006neocortical}, and not in a strongly-depolarized regime (\Fig{puzzle-fig}d-f). However, despite much progress (see Discussion for more details), a general theory of computation in such a regime has remained elusive, especially if computations are limited to smaller networks with only tens of neurons.

The second mismatched concept of rate networks concerns function approximation and its relationship to Dale's law, i.e, the common biologically-observed distinction between excitatory and inhibitory neurons \citep{eccles1976electrical}. The flexibility of rate networks relies upon the ability of each unit to linearly combine inputs in order to represent arbitrary input-output transformations on the network level (\Fig{puzzle-fig}c). To be most effective, the neurons' output rates must be combined with both positive and negative weights, resulting in mixed-sign connectivity on the level of individual neurons, and thus violating Dale's law. That said, several studies have successfully incorporated Dale's law (e.g., \cite{parisien2008solving, song2016training, miconi2017biologically, ingrosso2019training, shao2023relating}), and some have even suggested potential benefits to learning and robustness \citep{haber2022computational}. However, these studies have primarily taken a `bottom-up' approach, incorporating Dale's law for biological plausibility rather than for any computational necessity. It also remains to be seen how such sign constraints scale, as adding them to larger-scale machine learning benchmarks typically hinders (or at best matches) performance \citep{cornford2020learning, li2023learning}.

Neither of these mismatches is new and both have puzzled the field for a while. One common line of reasoning suggests that irregular firing and Dale's law are two among many manifestations of biological constraints on computation. From this view, one could justify the use of abstract rate networks as idealized versions of a constrained, noisy biological implementation. However, it is also possible that these mismatches reflect something fundamentally different about biological computation. Taking the latter perspective, here we offer a fresh approach to these puzzles. 

Our study is based on a combination of two recent developments: low-rank connectivity and spike-threshold geometry. Networks with low-rank connectivities generate low-dimensional dynamics in a latent activity space \citep{seung1996brain, eliasmith2003neural, landau2018coherent, mastrogiuseppe2018linking}. They thereby allow the activity of individual neurons to be linked to the modes or patterns of population activities often observed in real neural circuits \citep{gallego2017neural, saxena2019towards, keemink2019decoding, jazayeri2021interpreting, chung2021neural, langdon2023unifying}. Such insights can also be translated into spiking network models \citep{eliasmith2003neural, cimevsa2023geometry, koren2022biologically, depasquale2023centrality}.

However, low-rank connectivity alone does not guarantee spike-based computation in the balanced operating regime (e.g., \cite{eliasmith2003neural}). To achieve this, we take inspiration from spike-coding networks (SCNs) \citep{boerlin2013predictive}, whose function can be geometrically understood by visualizing spike-threshold boundaries in a space of latent population modes \citep{calaim2022geometry}. Importantly, this perspective places the boundary between sub- and supra-threshold voltages at the center of computation, leading to a fundamentally different operating regime from rate networks (\Fig{puzzle-fig}a,d). Such networks have been shown to exhibit irregular activity, and have several other desirable biological properties like robustness and energy efficiency \citep{deneve2016efficient,barrett2016optimal}. Based on a geometric reframing \citep{calaim2022geometry, mancoo2020understanding}, we generalize these networks here, removing several previous limitations of SCNs, while retaining their desirable properties.

We first show that a population of spiking neurons with rank-1 connectivity induces a latent variable readout and a spiking boundary in input-output space. We distinguish excitatory (E) and inhibitory (I) boundaries in this space, and show that a combined rank-2 EI network is a universal function approximator of static input-output transformations. Importantly, as we demonstrate, flexible function approximation can already be achieved in small networks, and can be theoretically understood without invoking large-network-size limits (e.g., mean field). Next, we illustrate a fundamental link between noise suppression and irregular firing, and demonstrate how mistuned connectivity between the two populations in the rank-2 EI network can lead to amplification of noise. We then consider effects of slower synaptic dynamics and transition delays on coding performance, and finally, we show that the respective low-rank spiking networks can be approximated by equivalent low-rank {\em rate} networks.

While we limit ourselves here to static function approximation in rank-1 populations with few neurons, in the Discussion we touch upon the implications for scaling up this framework to higher-rank networks with richer dynamical motifs, thereby providing a promising path to understand and construct spiking networks in biologically-realistic regimes.

%%%%%%%%%%%%%%%%%%%%%%%%%%%%%%%%%%%%%%%%%%%%%%%%%%%%%
%%%%%%%%%%%%%%%%%%%%%%%%%%%%%%%%%%%%%%%%%%%%%%%%%%%%%
\section{Spiking thresholds form convex boundaries in latent space}
\label{rank1section}

For simplicity, we will focus on networks with rank-one populations, and use them to develop the central concepts of this paper: latent variables and convex boundaries. Following an introduction of the general rank-1 case, we will distinguish inhibitory (I) and excitatory (E) populations. Then, we will combine the two populations into a rank-two EI network, and illustrate the resultant dynamics and input-output transformations.

\subsection{Rank-1 connectivity generates a latent variable} 

Let us consider a network of $N$ recurrently connected spiking neurons. We denote each neuron's membrane potential by $V_i$ and assume that the neuron fires a spike when it reaches a threshold, $T$. Each neuron's membrane potential will be described as a leaky integrate-and-fire neuron,
\begin{equation}
\tau\dot{V}_i(t) = -V_i(t) + F_i c(t) + \sum_{j=1}^N W_{ij} s_j(t) + b_i.\label{IODE}
\end{equation}
Here the first term on the right-hand-side describes a leak current, the second term a feedforward input, $c(t)$, weighted by synaptic feedforward weights, $F_i$, the third term the recurrent spike trains, $s_j(t)$, of presynaptic neurons, each weighted by recurrent synaptic weights, $W_{ij}$ (without sign constraints for now), and the last term is a constant background input. The constant background simply shifts the effective threshold of the neuron, and can therefore be modeled equivalently by setting $b_i=0$ and assuming an effective neural threshold $T_i$. For simplicity, we also set the neuron's membrane time constant $\tau=1$, so that the unit of time corresponds to $\tau$ (set to $10$ ms for all simulations).

We model each neuron's spike trains as a series of delta functions, such that $s_i(t) = \sum_f \delta(t-t_i^f)$, where $t_i^f$ is the time of the $f$-th spike of the $i$-th neuron. Importantly, for now we assume instantaneous communication between neurons without temporal synaptic dynamics or delays (this will be relaxed in Section 4; see Appendix). We furthermore assume that the diagonal of the weight matrix, $W_{ii}$, includes the voltage reset after a spike, thereby setting the reset voltage to
\begin{equation}
V_{i,\text{reset}} = T_i - W_{ii}.\label{reset}
\end{equation}

For our exposition, we will mostly rely on the integrated version of these equations \citep{gerstner2014neuronal, calaim2022geometry}. Let us first define the filtered input, $x(t)$, and the filtered spike trains, $r_i(t)$, as 
\begin{align}
\dot{x}(t) &= -x(t) + c(t) \label{filteredinput}\\
\dot{r}_i(t) &= -r_i(t) + s_i(t). \label{filteredspikes}
\end{align}
These equations express a filtering or convolution of the input or spike trains, respectively, with a one-sided exponential kernel, $h(t) = H(t)\exp(-t)$, where $H(t)$ is the Heaviside function (i.e., $H(t)=1$ if $t\geq 0$, and $H(t)=0$ otherwise). Accordingly, we can think of $r_i(t)$ as a simple model of a neuron's postsynaptic potential, or as a (single-trial) estimate of a neuron's instantaneous firing rate.

With these definitions, we can integrate \Eq{IODE} to obtain
\begin{equation}  
  V_i = F_i x + \sum_{j=1}^N W_{ij}r_j, \label{IODEint}
\end{equation}
where we have dropped the time index for brevity. By taking the derivative of this equation, inserting the equations for the filtered inputs and spike trains, \Eqs{filteredinput}{filteredspikes}, and remembering that $T_i=T-b_i$, one retrieves \Eq{IODE}. A key assumption for this section will be that the weight matrix
has rank one, so that we can write 
\begin{equation}
W_{ij} = E_i D_j.
\end{equation}
where we will call the scalars $E_i$ the encoding weights and $D_j$ the decoding weights. With this definition, the integrated voltage equation, \Eq{IODEint}, becomes
\begin{align}  
  V_i &= F_i x + \sum_{j=1}^NE_iD_jr_j \\
  &= F_i x + E_i\left(\sum_{j=1}^ND_jr_j\right). \label{IVE}
\end{align}
The term inside the brackets is simply a linear combination of the filtered spike trains, independent of index $i$. We will denote it as
\begin{equation}  
y = \sum_{j=1}^N D_j r_j, \label{readouty}
\end{equation}
so that the voltage becomes
\begin{equation}
V_i = F_i x + E_i y.  \label{IV}
\end{equation}
Importantly, the variable $y$ fully controls the dynamics of the network, in that knowledge of $y$ (together with the input $x$) is sufficient to compute the voltages, and consequently the spike trains. We will refer to $y$ as the readout, the output, or the {\it latent variable} of the network. 

We emphasize that linear, weighted sums of filtered spike trains are a common motif not only in the study of neural networks, but also in the analysis of population recordings \citep{fusi2016neurons, saxena2019towards, keemink2019decoding, vyas2020computation, jazayeri2021interpreting}. Indeed, such `linear readouts' are a standard means of extracting information from neural recordings, either through explicit linear decoding \citep{dayan2005theoretical} or through the use of linear dimensionality reduction methods such as principal components analysis or factor analysis \citep{cunningham2014dimensionality, pang2016dimensionality}. Consequently, we can also view $y$ as a component or mode of population activity.

\subsection{Inhibitory neurons form stable, attracting boundaries}\label{Isection}

So far we have not put any sign constraints on the connectivity. Now considering a population of inhibitory neurons with negative weights, we impose %$W_{ij}\leq0$ for all $i$ by
\begin{equation} 
W_{ij} = E_i D_j \leq 0\ \ \ \forall i,j, \label{Iconn}
\end{equation} 
by requiring
\begin{align}
    E_i\geq 0\ \ \forall i, \\
    D_j\leq 0\ \ \forall j.
\end{align}
With this sign convention, the negative sign is associated to the decoding weights, and from \Eq{readouty}, the latent variable or readout will also be constrained to be negative, $y\leq0$. The voltage equation, \Eq{IV}, remains unchanged, but the positive encoding weights $E_i$ and negative latent variable, $y$, ensure that each neuron receives negative feedback from the population.

Our first goal will be to understand the dynamics of the integrate-and-fire network in the joint space of inputs and readouts. Using \Eqs{filteredspikes}{readouty}, we can compute the derivative of the readout,
\begin{equation} 
\dot{y} = -y + \sum_{j=1}^N D_j s_j, \label{readoutdydt}
\end{equation} 
revealing the dynamics to be a weighted, leaky integration of the network's spike trains. These dynamics therefore fall into two regimes: 
\begin{align} 
  \dot{y} &= -y \nonumber  \\ 
  y &\leftarrow y + D_i \qquad\text{if}\qquad V_i=F_ix + E_iy \geq  
      T_i. \label{Iydyn}
\end{align} 
Either the readout is leaking towards zero (top equation) or it is changing abruptly due to spikes (bottom equation).

\begin{figure}[t!]
\centering
  \includegraphics[scale=1]{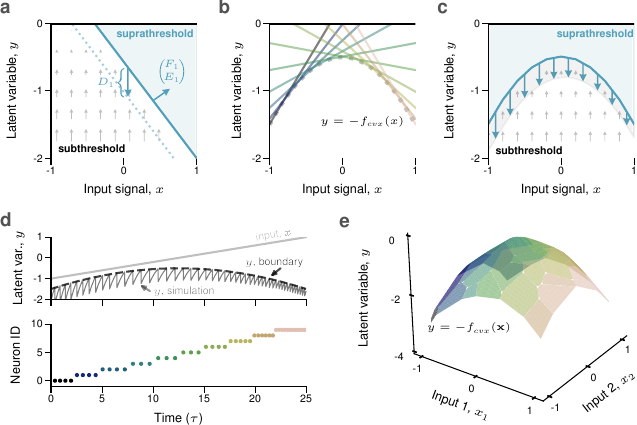}
  \caption{\label{Ibounds1}
    A rank-1 inhibitory population forms a stable, attracting boundary. 
    \textbf{a}: A single inhibitory neuron's voltage (\Eq{IV}) is visualized in input-output space; the spike threshold ($V_1=T_1$; solid blue line, with normal vector $(F_1, E_1)$) divides the space into sub- and supra-threshold sets (white and blue shaded area). Grey arrows illustrate dynamics due to the leak; downward facing blue arrow illustrates the effect of a spike at the boundary, which sets $y \rightarrow y+D_1$ (dotted blue line).
    \textbf{b}: The thresholds of multiple ($N=10$; colored lines) neurons were tuned to delineate a concave (or negative convex) boundary, $y=-\fcvx(x)=-x^2-\tfrac{1}{2}$ (dotted gray line).
    \textbf{c}: The boundary (blue line) divides the input-output
    space into a convex subthreshold set (with {\em all} voltages
    below threshold), and a suprathreshold set (with {\em at least one}
    voltage above threshold), forming a stable input-output relationship. Output precision is defined as the distance between the neural thresholds and the resets after a spike, which can be different for each neuron (gray shading; see \Eq{precision}).
    \textbf{d}: Latent variable output (top) and spike raster (bottom) of a simulation of the spiking network from panel \textbf{c} (colors as in \textbf{b}) for a time-varying input from $x=-1$ to $1$ (light gray, top; time in units of $\tau$); the spiking simulation output (solid dark gray) follows the true boundary (dashed black line), and each input value is coded by a single neuron.
    \textbf{e}: A negative convex surface boundary for a two-dimensional input (\Eq{vboundND}), made up of 36 neurons (colored segments).
  }
  \end{figure}

To gain more intuition for these separate regimes, let us start with a single neuron. For this neuron, the equation $V_1=F_1x+E_1y=T_1$ defines a line in $(x,y)$-space, i.e., in the space of inputs and outputs (\Fig{Ibounds1}a, solid blue). On one side of the line, the neuron is subthreshold, and on the other side, it is suprathreshold. The output $y$ always leaks towards zero in the absence of firing. Once the neuron's threshold is hit, it fires, and the readout bounces down according to the size of the neuron's decoder weight, to $y \leftarrow y + D_1$. From a biophysical point of view, the neuron's inhibitory self-connection acts as a reset, and the voltage moves from threshold ($V_1=T_1$) to a hyperpolarized value.

Now considering a network of several neurons, each threshold will trace a different line in the input-output space (\Fig{Ibounds1}b) and together they will form a single boundary (\Fig{Ibounds1}b, gray dashed line; \Fig{Ibounds1}c, solid blue line). We will use `subthreshold' to denote the region where \emph{all} neurons' voltages are below threshold and `suprathreshold' to denote the region where \emph{at least} one neuron is above threshold. Since $E_i\geq 0$, the subthreshold region lies below each neuron's threshold line. Just as in the single-neuron case, the latent variable, $y$, leaks towards zero until it reaches the boundary, where a neuron will fire a spike, causing $y$ to jump back into the subthreshold region, and the dynamics continue. Note that this jump into the subthreshold region pushes $y$ away from all threshold boundaries, and therefore corresponds to recurrent inhibition between all neurons. Since each part of the boundary is covered by a single neuron, only one neuron will fire spikes for each value of the input $x$ (\Fig{Ibounds1}d).

We emphasize that \emph{any} integrate-and-fire network following \Eq{IODE} and assuming the sign and rank-constraints, $W_{ij}=E_iD_j\leq 0$, will form a stable spike-threshold boundary. However, a rank-1 population with randomly-distributed parameters, $(F_i, E_i, T_i)$, will, more often than not, yield a boundary composed of only a subset of neurons (see Appendix \ref{appendix:random} for more details). In other words, some neurons' boundaries will be well within the supra-threshold set, and so will never fire any spikes. Furthermore, a random boundary may also cross the $x$-axis ($y=0$), resulting in some input values for which no neuron is active. Since such parameter regimes include silent neurons or fully silent activity regimes, we consider them degenerate and do not consider them. Instead, the networks shown in \Fig{Ibounds1}b,c,e were precisely tuned so that each neuron's threshold lies along a smooth quadratic function (see Appendix \ref{appendix:Inh-params}).

\subsection{The boundary determines the input-output mapping}

The concept of a boundary in the input-output space will be central to our developments. This perspective can also be generalized to rank-1 networks with multiple inputs. Let us write $\bx=(x_1,x_2,\ldots, x_M)$ for an $M$-dimensional input, in which case the voltage equation becomes 
\begin{equation}  
V_i = \bF_i^\top \bx + E_i y \leq T_i. \label{vboundND}
\end{equation}
This inequality still describes a linear boundary, but now in a higher-dimensional input-output space, $(\bx,y)\in\mathbb{R}^{M+1}$. Multiple neurons again define different boundaries, each of which divides the input-output space into two half-spaces. The intersection of the individual subthreshold half-spaces yields the
population's subthreshold set, which is guaranteed to be convex. The boundary of this convex set can therefore be written as
\begin{equation}
 y = -\fcvx(\bx),
\end{equation}
where $\fcvx(\bx)$ denotes a (piecewise-linear) convex function. The negative sign implies that $y$ is actually a  {\em concave} function of the inputs $x$. For $\bx\in\mathbb{R}^2$, we plot an example boundary in \Fig{Ibounds1}e.

The above function serves as an idealized description of the network's input-output relationship. More precisely, the output of the network jumps back and forth between this threshold boundary and the set of $y$-values that is reached after a spike (\Fig{Ibounds1}d). Since every input $\bx$ can be uniquely associated with the neuron whose boundary is exposed, we can define a `decoder' function, $\mathcal{D}(\bx)\leq 0$, which takes the value of the decoder for the neuron that becomes active for input $x$. Accordingly, the output takes values in the interval 
\begin{equation}
y=[-\fcvx(\bx)+\mathcal{D}(\bx),-\fcvx(\bx)]. \label{precision}
\end{equation}
We refer to this deviation from the ``true'' boundary, $-\fcvx(\bx)$, caused by discrete spiking events, as the \emph{precision} by which the input is mapped onto the output, as determined by the function $\mathcal{D}(\bx)$ (compare \Fig{Ibounds1}c, gray shading and \Fig{Ibounds1}d).

\subsection{The boundary determines the latent dynamics}

We can write down the dynamics of $y$ in two equivalent ways. First, we can think of each individual neuron as enforcing its own threshold, and we can think of the boundary as the joint action of all thresholds. Merging the two dynamical regimes, \Eq{Iydyn}, into a single equation, we obtain
\begin{equation} 
\dot{y} = -y + \sum_{i=1}^N D_i \, I\big( \bF^\top_i\bx + E_iy - T_i\big),\label{yIneurons}
\end{equation}
where the indicator function $I(\cdot)$ denotes an infinitely high boundary, i.e., $I(z) = \infty$ if $z\geq 0$ and $I(z)=0$ otherwise. The indicator function is commonly used in convex analysis \citep{rockafellar1970convex, boyd2004convex}, and can be understood similarly to the delta-function used in spiking networks, i.e., as a limiting case, $I(z)=\lim_{\Delta t\rightarrow 0} H(z)/\Delta t$, where $H(z)$ is the Heaviside function, and $\Delta t$ the integration time step.

Second, we can think of the network as forming a single (globally) stable boundary
\begin{equation} 
\dot{y} = -y + \mathcal{D}(\bx) I\left( B(\bx,y) \right).  \label{yIboundary}
\end{equation}  
using the decoder function $\mathcal{D}(\bx)$ and similarly defining a boundary function
\begin{equation}
B(\bx,y) = y + \fcvx(\bx),
\end{equation}
where we can recover our definition of the boundary input-output function by setting $B(\bx,y)=0$. \Eqs{yIneurons}{yIboundary} are mathematically equivalent, and both will prove useful later on.

\subsection{Excitatory neurons form unstable, repellent boundaries}\label{Esection}

Next, we consider a network of $N$ recurrently connected, excitatory neurons. Much of our exposition follows the same outline as for the inhibitory network, and we will mostly focus on highlighting the differences. Of course, the key difference is positive connectivity, compared to \Eq{Iconn},
\begin{equation}
W_{ij} = E_iD_j \geq 0,
\end{equation}
with constraints on the encoders and decoders
\begin{align}
E_i\geq 0\ \ \forall i,\\
D_j \geq 0\ \ \forall j. 
\end{align}
Given these sign conventions for the excitatory network, the latent variable or readout will be constrained to be positive, $y\geq0$. The voltage equation, \Eq{IV}, again remains the same, but now each neuron receives positive feedback from itself and all other neurons of the excitatory population.

We note that due to the positivity constraint, even the diagonal terms of the connectivity matrix, $W_{ii}$, are all positive. For the inhibitory network, in contrast, the diagonal terms were negative and thereby corresponded to the self-reset current after a spike. A positive diagonal term here means that the neuron will self-excite, and thereby immediately fire more spikes. While the absence of a reset term may seem unnatural, it provides a useful entry point for studying low-rank, all-excitatory networks in their idealized form. (In practice, a self-reset can be introduced, which we describe below).

\begin{figure}[t!]
\centering
    \includegraphics[scale=1]{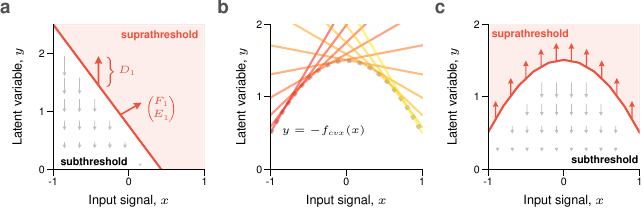}
  \caption{\label{Ebounds1}
   A rank-1 excitatory population forms an unstable, repellent boundary. 
   \textbf{a}: Input-output space for a single excitatory neuron (compare with and refer to caption of \Fig{Ibounds1}a). Note that a spike moves the latent variable further into the supra-threshold set, and the leak dynamics decay away from the boundary.
   \textbf{b}: The thresholds of multiple ($N=10$, colored lines) excitatory neurons were tuned to delineate a concave (or negative convex) boundary in input-output space, $y=-f_{cvx}(x) = -x^2 + \tfrac{3}{2}$ (dotted gray line).
   \textbf{c}: The boundary forms a convex subthreshold set (where {\em all} voltages
    are below threshold), analogous to the inhibitory population (Fig.~1). Unlike the inhibitory population, however, the boundary itself is now unstable --- it either leaks to zero or explodes.}
\end{figure}

The dynamics of an example network are visualized in \Fig{Ebounds1}. Once more, each neuron divides the input-output space into two halves, and the boundary of these half spaces is given by equating the voltage equation (\Eq{IV}) with the threshold, i.e., $V_i=T_i$ (\Fig{Ebounds1}a). The sub- and suprathreshold regions are similarly defined as in the all-inhibitory network and are shown for a single neuron or several neurons in \Fig{Ebounds1}a,b. The output $y$ is now positive, and still leaks towards zero in the subthreshold region, but now this leak is \emph{away} from the boundary. If the boundary is breached, however, the respective neuron fires, which moves the latent variable further into the supra-threshold regime. The spike causes self-excitation of the firing neuron, and, potentially, the crossing of other neurons' thresholds. (Strictly speaking, this regime is mathematically ill-defined. For now, we will be pragmatic, discretize time, and assume that only one neuron can fire per time step. In Section~\ref{ratesection}, we will relax this assumption.) In consequence, the dynamics beyond the boundary self-reinforces the growth of the output and becomes highly explosive (\Fig{Ebounds1}c). To illustrate the unstable dynamics on either side of the boundary, the boundary function in \Fig{Ebounds1}b,c was chosen such that $y=0$ is contained within the subthreshold set for the displayed values of $x$, meaning that in the absence of previous spiking, the network will remain silent. This choice is primarily illustrative, and we will return to the spontaneously-active case in the next section after re-introducing inhibition.

Just as in the inhibitory network, the set of subthreshold readouts is a convex set, and the boundary is a concave function of the input within a given range (see \Fig{Ebounds1}). The same reasoning holds when the input $\bx$ becomes multi-dimensional. We can therefore characterize the unstable boundary through a concave function,
\begin{equation}
y = -\fcvx(\bx),
\end{equation}
defined on the set of inputs, $\bx$, for which $y\geq 0$. Since this boundary is unstable, it does not per se describe the input-output mapping of an all-excitatory network. Indeed, the input-output mapping of the network will be determined by the mechanism(s) that stabilize the explosion (see below). Finally, the dynamics of the excitatory network also follow \Eqs{yIneurons}{yIboundary}, with the key difference that $D_i$ or $\mathcal{D}(\bx)$ are now positive, which inverts the dynamics around the boundary.

Given that the excitatory population is unstable, some other mechanisms needs to kick in to stabilize the dynamics. One stabilizing factor is the leak of each neuron, which eventually could counter the explosion of activity. A second approach would be to give each excitatory neuron a self reset and/or refractory period after firing a spike. This can be formalized by adding a diagonal component $\mu$ to the connectivity matrix such that
\begin{equation}
    W_{ii} = E_iD_i - \mu,\label{mu-eqn}
\end{equation}
which serves as a ``soft'' refractory period. Though technically this breaks the assumed low-rank connectivity structure, it will prove useful, and we will return to it in Section 3. Finally, a third possibility is to stabilize the network activity with inhibition, which we will do next.

Before moving on, we make one additional comment about Dale's law here. The concepts of stable and unstable spike-threshold boundaries form the main computational components of our framework. Intriguingly, sign-constrained inhibitory and excitatory populations are sufficient to generate such well-behaved boundaries, and we will see below how the combination of the two leads to a unique computational regime. We note that while a stable boundary can be generated without adhering to Dale's law, as is known from previous work on spike-coding networks \citep{calaim2022geometry}, the stability is only local and can rapidly explode upon certain perturbations (see Discussion).

\subsection{The inhibitory boundary can stabilize the excitatory boundary}

\label{EIsection}
We now study networks of coupled excitatory (E) and inhibitory (I) neurons. In order to do so, we will first go back to differential equations. Assuming that there are $N^I$ inhibitory and $N^E$ excitatory neurons, we again use leaky integrate-and-fire neurons to describe the membrane potentials of the two populations,
\begin{align} \dot{V}_i^I &= -V^I_i + F_i^I c(t)
                + \sum_{j=1}^{N^E} W^{IE}_{ij}s^E_j(t) + \sum_{j=1}^{N^I}
                W^{II}_{ij}s^I_j(t) \\
                \dot{V}_i^E &= -V^E_i + F_i^E c(t)
                              + \sum_{j=1}^{N^E} W^{EE}_{ij}s^E_j(t) +
                              \sum_{j=1}^{N^I} W^{EI}_{ij}s^I_j(t).
              \label{EIODE}
\end{align}
Besides the self-connections within the subnetworks, $W^{EE}$ and $W^{II}$, we also introduce cross-connections between the two networks, designated by the matrices $W^{EI}$ and $W^{IE}$. The thresholds of the two populations are given by $T_i^I$ and $T_i^E$, and as before, may incorporate possible background inputs.

Just as before, we assume that the self-connection matrices are rank-1. We will furthermore assume that the cross-connection matrices are likewise rank-1, while sharing the same decoders. Specifically, we set (for all $i,j$)
\begin{align} 
W^{II}_{ij} &= E^{II}_i D^I_j \leq 0 \label{EI_WII}\\ 
  W^{IE}_{ij} &= E^{IE}_i D^E_j \geq 0\label{EI_WIE}\\
  W^{EI}_{ij} &= E^{EI}_i D^I_j \leq 0 \label{EI_WEI}\\ 
W^{EE}_{ij} &= E^{EE}_i D^E_j \geq 0, \label{EI_WEE} 
\end{align} 
resulting in a rank-2 excitatory-inhibitory (EI) network. These settings are not fully general, as the cross-connection matrices could, in principle, have their own set of decoders. We will defer the analysis of the general case to Section~\ref{noisecontrol}. All the encoding weights are assumed to be positive, while the decoders follow previously-specified sign constraints, i.e., $D^I_j\leq 0$ and $D^E_j\geq 0$, ensuring Dale's law.

Given the above assumptions, we define two latent variable readouts,
\begin{align}
y^I &= \sum_{j=1}^{N^I} D^I_j r^I_j\\
y^E &= \sum_{j=1}^{N^E} D^E_j r^E_j.
\end{align}
which allows us to integrate the differential equations and obtain
\begin{align}
V^I_i &= F_i^I x + E_i^{IE} y^E + E_i^{II}y^I \leq T_i^I\label{EInet_VI}\\
V^E_i &= F_i^E x + E_i^{EE} y^E + E_i^{EI}y^I \leq T_i^E.\label{EInet_VE}
\end{align}
We see that even in this more complicated network, each neuron can once more be interpreted as a bound in input-output space. The key difference is that we now have a rank-2 network, and the input-output space has become three-dimensional, given by $(x,y^E,y^I)\in \mathbb{R}^3$. As a consequence, neural thresholds have become planes instead of lines. Furthermore, we now have two distinct sets of spike-threshold boundaries, corresponding to the two populations. 

The spiking dynamics of the network can again be understood by focusing on the space of latent variables, which is now two-dimensional. Taking the derivative of the readout equations, we obtain the two-dimensional dynamics of the latent space, compare \Eq{yIneurons},
\begin{align} 
\dot{y}^I &= -y^I + \sum_{i=1}^{N^I} D^I_i \, I\big( F^I_i x + E^{IE}_iy^E 
  + E^{II}_iy^I - T^I_i\big) \label{Idyn_EI}\\
\dot{y}^E &= -y^E + \sum_{i=1}^{N^E} D^E_i \, I\big( F^E_i x + E^{EE}_iy^E 
  + E^{EI}_iy^I - T^E_i\big). \label{Edyn_EI}
\end{align} 
We have already done all the work to understand these equations. In the first equation, for instance, we can simply understand $y^E$ as a (time-varying) external input. As a consequence, we can treat the inhibitory population as receiving a two-dimensional input, $(x,y^E)$ (see \Fig{Ibounds1}e), so that the derivations from Section~\ref{Isection} all remain the same. Similarly, the excitatory population can be treated as receiving a two-dimensional input $(x,y^I)$. Accordingly, the inhibitory dynamics generates a stable boundary in $(x,y^E,y^I)$-space, and the excitatory dynamics generates an unstable boundary in $(x,y^E,y^I)$-space. 

\begin{figure}[t!]
\centering
  \includegraphics[scale=1]{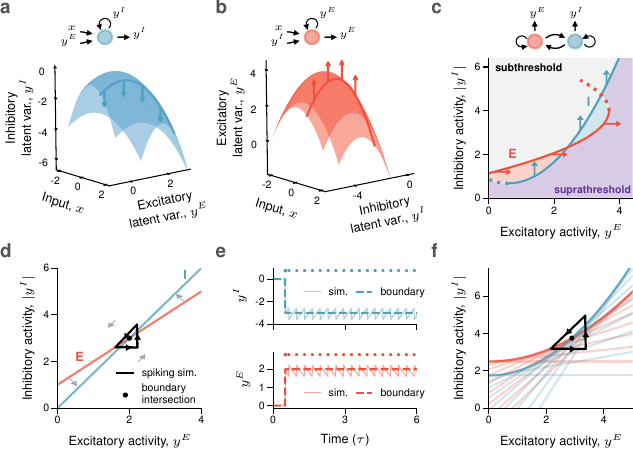}
  \caption{\label{EI_bounds}
    The inhibitory boundary can stabilize the excitatory boundary. 
    \textbf{a}: The inhibitory population, with latent readout $y^I$, forms a negative convex, attracting boundary in $(x,y^E)$ space.
    \textbf{b}: The excitatory population, with latent readout $y^E$, forms a negative convex, repellant boundary in $(x,y^I)$ space.
    \textbf{c}: For fixed $x$ (blue and red lines in panels \textbf{a} and \textbf{b}, respectively, for $x=4$), the inhibitory and excitatory boundaries can be viewed in $(y^E,|y^I|)$ activity space, and delineate four distinct regions of the state space: $E$ and $I$ subthreshold (gray), $E$ suprathreshold and $I$ subthreshold (red), $E$ subthreshold and $I$ suprathreshold (blue), and both $E$ and $I$ suprathreshold (purple).
    \textbf{d,e}: Single-neuron E and I boundaries with spiking simulation dynamics (black trajectory in panel \textbf{d}; blue and red solid lines in \textbf{e} (sim.)) around the boundary crossing (black dot in \textbf{d}; blue and red dashed lines in \textbf{e}).
    Dots in panel \textbf{e} show the spikes fired.
    \textbf{f}: Heterogeneous E and I boundaries, in which I (respectively E) neurons (faded blue and red lines) form a piecewise-linear approximation to the idealized convex boundaries (opaque blue and red lines). Dynamics are shown in black, oscillating around the boundary crossing (black dot).
  }
  \end{figure}

Example boundaries for the inhibitory and excitatory populations are shown in \Fig{EI_bounds}a,b, respectively. The blue boundary describes the inhibitory population, and is a negative convex function of $x$ and $y^E$, denoted by $y^I = -\fcvxI(x,y^E)$. While we illustrate the boundary as a smooth continuous surface, in reality it will be piecewise linear, made up of individual neurons' thresholds as in \Fig{Ibounds1}e (see Appendix \ref{appendix:EI-params}). Since all decoders are negative, each spike of an inhibitory neuron drives the readout into the subthreshold regime, and the state space is restricted to negative values of $y^I$ (\Fig{EI_bounds}a). Similarly, the excitatory population is described by the red, negative convex boundary (\Fig{EI_bounds}b), which is a function of $x$ and $y^I$, denoted $y^E = -\fcvxE(x,y^I)$. However, in this case each excitatory spike drives the readout into the suprathreshold regime towards positive values of $y^E$.

Following \Eq{yIboundary}, we can also rewrite the dynamics in terms of the boundaries only,
\begin{align} 
\dot{y}^I &= -y^I + \mathcal{D}^I(x,y^E) I\big( B^I(x,y^E,y^I)\big) \\
\dot{y}^E &= -y^E + \mathcal{D}^E(x,y^I) I\big( B^E(x,y^E,y^I)\big) 
\end{align} 
where the boundary functions are given by $B^I(x,y^E,y^I)=\fcvxI(x,y^E) + y^I$ and $B^E(x,y^E,y^I)=\fcvxE(x,y^I) + y^E$, and the boundaries are retrieved by setting them to zero. To visualize the dynamics of this joint system, we assume a constant input of $x=4$, which means that the two boundary surfaces reduce to 1-d curves in $(y^E,y^I)$-space (shown as the blue and red lines in \Fig{EI_bounds}a,b, respectively). Both 1d-curves are plotted together in \Fig{EI_bounds}c. Here, we plot the absolute value of the inhibitory readout, $|y^I|$, as it is comparable to the summed population activity (especially if decoders are constant, $\mathcal{D}^I(x,y^E)=\text{const}$), and thereby relates to typical ways of looking at EI networks \citep{wilson1972excitatory, dayan2005theoretical, gerstner2014neuronal}.

We see that the two boundaries intersect (\Fig{EI_bounds}c), and we can distinguish four regions of the state-space around this intersection. If both populations are below threshold, the population activities will simply leak toward zero (gray area in \Fig{EI_bounds}c). If one population is suprathreshold and the other subthreshold (red or blue areas in \Fig{EI_bounds}c), then the suprathreshold population activity increases rapidly (through spiking) while the subthreshold activity decreases slowly (through the leak). If both populations are suprathreshold, both activities increase rapidly (purple areas in \Fig{EI_bounds}c). In order to eliminate any ambiguities in the suprathreshold regime, and to promote stability, we will prescribe that inhibitory neurons always fire before excitatory neurons when both are above threshold. This somewhat ad hoc rule implies that inhibition generally has faster dynamics than excitation, and will be relaxed later in Section~4.

The latent dynamics of the EI system in \Fig{EI_bounds}c will tend to gravitate around the point at which the two boundaries cross. The dynamics resemble a stable limit cycle, which, in the limit as the decoders become smaller and smaller, effectively becomes a stable fixed point for certain requirements of the two boundaries. Intuitively, we can understand the stability of the EI system by considering the slopes of the two boundaries in ($y^E,|y^I|)$-space (\Fig{EI_bounds}d). Under the rule that inhibitory neurons always fire before excitatory neurons, local stability requires the inhibitory boundary to have a steeper slope, such that it is able to push the dynamics into the subthreshold area when the readouts are above the crossing point. For instance, if the network is composed of one inhibitory neuron and one excitatory neuron (\Fig{EI_bounds}d), this slope condition can be succinctly expressed through the encoding weights as
\begin{equation}
\frac{E^{IE}_1}{E^{II}_1} > \frac{E^{EE}_1}{E^{EI}_1},\label{EI-stability}
\end{equation}
where both neurons have been labeled with index $i=1$. This inequality is analogous to the condition on the determinant in a two-dimensional linear stability analysis \citep{izhikevich2007dynamical}. A more formal treatment of stability can be done after relating the spiking dynamics to those of a rate network (Section 4; \citep{dayan2005theoretical, izhikevich2007dynamical}). We also note that other stable dynamical regimes may be present in the more general case, such as networks in which the excitatory or both populations are silent. As before, we consider these to be degenerate and do not discuss them here.

To illustrate the stable EI dynamics, we simulate the two-neuron system in \Fig{EI_bounds}d,e. Here, we have a single excitatory and a single inhibitory neuron. When the two outputs $y^I$ and $y^E$ leak towards zero, both neurons experience a decrease in recurrent inhibition and recurrent excitation, resulting in an overall depolarization of both neurons. The excitatory neuron fires first, and the output $y^E$ moves to the right. At this point, both neurons experience an excitatory postsynaptic current. The inhibitory neuron then crosses threshold and, according to our rule above, fires before the excitatory neuron can fire another spike. The inhibitory spike inhibits the excitatory neuron and moves the output $(y^E,|y^I|)$ into the subthreshold set of both neurons. At this point, the dynamics repeat. While globally stable, the local dynamics of the limit cycle, and repeatability of the spiking pattern, not only depends on the shapes of the boundaries, but also the decoding weights (see Appendix \ref{appendix:EI-params}). Principally the same pattern is observed when we consider a network with several neurons in each population (\Fig{EI_bounds}f). In the deterministic regime that we consider in this Section, this oscillatory pattern only encompasses the two neurons that make up the crossing point for the given input $x$ (and local stability can still be assessed with \Eq{EI-stability}), but the latent dynamics are reflected in all neurons' voltages (not shown).

We note that in the general case, with multiple neurons per population, the E and I boundaries can in principle cross multiple times, leading to more interesting dynamics such as bistability (not shown, but see Discussion).

\subsection{The rank-2 EI network can approximate arbitrary, non-linear input-output functions}

\begin{figure}[t!]
\centering
      \includegraphics[scale=1]{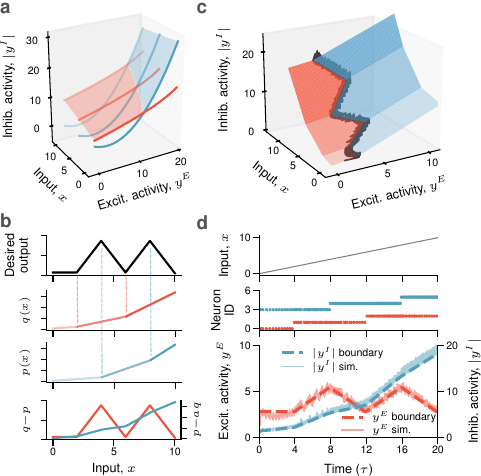}
      \caption{\label{EI-function-approx} Universal function approximation in the rank-2 EI network. \textbf{a}: Each of three input levels (illustrated for $x=0$, $5$, and $10$) results in a different intersection of the excitatory (red lines) and inhibitory (blue lines) population boundaries, thereby modulating the latent outputs of the network. Note that as $x$ changes, the pair of E and I neurons that form the boundary intersection may change (not shown, but see panel \textbf{c}). The red (or respectively blue) shading represents the area for which the excitatory (respectively inhibitory) boundary is above the other with respect to $|y^I|$, highlighting the changing boundary intersection as a function of $x$. \textbf{b}: A desired saw-like output function (top) can be decomposed into the difference of two convex functions $q(x)$ and $p(x)$ (middle), such that the excitatory latent variable approximates the function (red curve, bottom). \textbf{c}: Visualization of a network of excitatory ($N^E=3$ red planes) and inhibitory ($N^I=3$ blue planes) neurons that approximates the saw-like function following panel \textbf{b}. Note that the spiking dynamics (black faded line, see panel \textbf{d}) follows the boundary intersection as $x$ increases. \textbf{d}: Simulating the spiking network from \textbf{c} for an input $x$ from $0$ to $10$ (top) confirms that the excitatory (red) and inhibitory (blue) readouts follow the boundaries (bottom). The spike trains demonstrate how different pairs of E and I neurons code for the boundary at different values of $x$ (middle).}
  \end{figure}

As shown above, provided that stability conditions are satisfied, the dynamics of the rank-2 EI network will converge towards an oscillation around the crossing point of the two boundaries. Crucially, this crossing point depends on the input $x$. If we add this third dimension back to the picture, we see that the crossing point between the two populations can vary if different neurons form the boundary at different values of $x$ (\Fig{EI-function-approx}a). It turns out that such an approach can yield a rich set of possible input-output functions. Indeed, while the above equations cannot generally be solved for $y^I$ or $y^E$, we can show that each latent variable on its own can, in principle, be any function of the input $x$. 

We assume that the inhibitory population boundary has the form
\begin{equation}
    y^I = -\fcvxI(x,y^E)=-p(x)-ay^E,
\end{equation} 
where $p(x)$ is some convex function, and $a$ is a positive constant. Then, we assume the excitatory population boundary has the form 
\begin{equation}
    y^E = -\fcvxE(x,y^I) = -q(x) - y^I,
\end{equation}
where $q(x)$ is also a convex function. We note that the stability condition, \Eq{EI-stability}, requires $a>1$, which ensures that the inhibitory boundary has a larger slope than the excitatory boundary, and we will use $a=2$ here. Given that the input-output function will be described by the crossing of the two boundaries, we can rewrite the excitatory latent variable as
\begin{align}
  y^E &= -q(x) - y^I\\
      &= -q(x) + p(x) + 2y^E. 
\end{align}
Finally, rearranging terms, we obtain
\begin{align}
    y^E = q(x) - p(x),
\end{align}
which is the difference of two convex functions. Instead solving for $y^I$, we get
\begin{align}
    y^I = p(x) - 2q(x),
\end{align}
which is again the difference of two convex functions. Since any continuous function with a bounded second derivative can be expressed as the difference of two convex functions\citep{yuille2003concave}, the input-output function of the rank-2 E-I network is fully general (see Appendix \ref{appendix:diff-cvx} for more details). Accordingly, the limitations to computations imposed by the boundaries of the separate populations disappear once excitatory and inhibitory populations are coupled. Furthermore the interpretation of the input-output transformation as computing a difference of two convex functions puts forth an intriguing computational hypothesis for the function of Dale's law (see Discussion).

A toy example of function approximation is shown in \Fig{EI-function-approx}b-d, in which a non-convex, saw-like function is approximated by the excitatory latent readout of a small network with $N^E=3$ and $N^I=3$~neurons. Such a network computation is intuitively straightforward to visualize and to construct (see Appendix). Following the notation above, we can simply design piecewise-linear convex functions $p(x)$ and $q(x)$ such that their difference approximates (or equals, in this case) the desired function (\Fig{EI-function-approx}b). We note that more continuous functions can be approximated with more neurons, and similar non-convex approximations can also be done with the inhibitory readout (not shown).

While the above derivation was done for a one-dimensional input $x$, it also holds for $M$-dimensional inputs, so that the rank-2 EI network can approximate arbitrary mappings from $\mathbb{R}^M$ to $\mathbb{R}$. For simplicity, we limit ourselves here to rank-2 networks and 1d outputs, which also results in stereotypical, non-overlapping spike trains (\Fig{EI-function-approx}d). We contend that this is largely due to the simplicity of the 1d task, which should be remedied when considering higher-rank networks (see Discussion).

%%%%%%%%%%%%%%%%%%%%%%%%%%%%%%%%%%%%%%%%%%%%%%%%%%%%%
%%%%%%%%%%%%%%%%%%%%%%%%%%%%%%%%%%%%%%%%%%%%%%%%%%%%%
\section{Synaptic connections can suppress or amplify noise}\label{noisecontrol}

Thus far, we have studied deterministic networks, and we have limited ourselves to rank-2 EI networks in which both populations share the same decoders (see \Eq{EI_WII}-\Eq{EI_WEE}), i.e., excitatory neurons see the same inhibitory latent variable $y^I$ as the inhibitory neurons themselves (and the same for the excitatory latent, $y^E$). Here, we will show how relaxing this constraint and adding small amounts of noise to each neuron leads to networks that can control or amplify input-independent noise.

\subsection{Noise makes boundaries jitter}

\begin{figure}[t!]
\centering
  \includegraphics[scale=1]{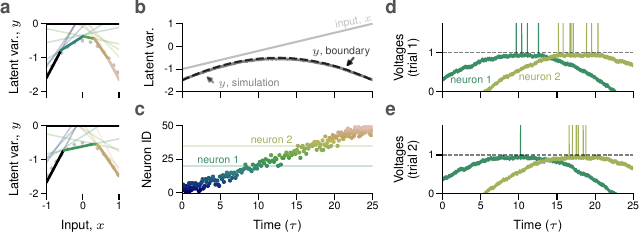}
  \caption{\label{simulnoise} Noise causes jittery boundaries and irregular firing in the inhibitory population. \textbf{a}: Current noise makes individual neural threshold boundaries (faded colored lines) jitter up and down independently, resulting in a fluctuating population boundary (opaque multi-colored line) around the default boundary function (gray dotted line). Two different realizations of noise are shown (top and bottom) and illustrate that each input value may be coded by a different neuron at different times. \textbf{b}: Input $x$ (light gray) and latent readout $y$ (dark gray) for a spiking simulation of a network with $N=50$ neurons, with a small amount of independent current noise added. Default boundary is shown in dashed black (compare with panel \textbf{a}, dotted gray). \textbf{c}: Spike trains for the simulation in (\textbf{b}); note that several neurons become active for any given input $x$. \textbf{d,e}: Voltages of two example neurons (highlighted in panel \textbf{c} for trial 1) for two trials from the network in (\textbf{b,c}). Note that the spike trains are variable despite the high precision of the output $y$.}
\end{figure}

To study the addition of noise, we will begin by returning to the single, inhibitory population. Concretely, we assume that all neurons are subject to small independent white noise currents, $\eta_i(t)$, which we add to the differential equation, \Eq{IODE}, of each neuron. This white noise will be filtered through the leaky integration of the membrane, so that the integrated voltage, 
\Eq{IVE}, becomes 
\begin{equation}  
V_i = F_ix + E_iy  + h*\eta(t) \; \leq \; T_i,
\label{Voltage}
\end{equation}  
where $h(\cdot)$ describes a one-sided exponential kernel, and ``$*$'' denotes a convolution. Following \cite{calaim2022geometry}, we move the noise term onto the threshold,
\begin{equation} 
F_ix + E_iy  \; \leq \; T_i - h*\eta(t),
\end{equation} 
and re-interpret this equation geometrically. From this view, the white noise causes random shifts in the precise position (but not the orientation) of a neuron's boundary in input-output space (\Fig{simulnoise}a). Additionally, and again following \cite{calaim2022geometry}, we now consider neurons to have an additional self reset, following \Eq{mu-eqn}. This acts as a soft refractory period, and can be modeled as an additional threshold term
\begin{equation} 
F_ix + E_iy  \; \leq \; T_i - h*\eta(t) + \mu r_i(t),\label{NoisyThreshVoltage}
\end{equation} 
which increases the neuron's threshold after spiking (recall that $r_i(t)$ is the filtered spike train, see \Eq{filteredspikes}). This can be considered as another mechanism that causes temporary shifts in the threshold of each neuron --- in this case, the threshold will jump away from the population boundary by a fixed value after spiking, and exponentially decay back to its original location. At the network level, due to both of these effects, we can thus see that the boundaries of all neurons will jitter around their default positions, as illustrated for two noise snapshots in \Fig{simulnoise}a, resulting in a noisy population boundary.

\subsection{A jittery boundary causes irregular spiking}

The key consequence of the jittery boundary is that more than one neuron can fire for a given input $x$. In the fully deterministic, inhibitory network, we saw that only one neuron becomes active for any given input $x$ (compare \Fig{Ibounds1}d). However, once we add noise to the network, we observe instead that several neurons may become active for any input value $x$, even though the latent variable is still faithfully represented (\Fig{simulnoise}b,c).

Geometrically, the individual thresholds start fluctuating around their default position, and if a subset of neural thresholds are close to each other, then the noise and self-reset can push any of these thresholds to the front. As a consequence, neurons within this subset take turns in enforcing the boundary for any particular input value $x$. When the output $y$ decays towards the boundary, it reaches the threshold of the neuron that just happens to be more sensitive at that time point.

This type of redundancy can lead to neurons firing in a seemingly random fashion, simply because the latent variable $y$ can reach one of several thresholds, depending on the noise in the system. \Fig{simulnoise}d,e shows two trials of the inhibitory network. Just as in previous SCN work \citep{boerlin2013predictive, deneve2016efficient, calaim2022geometry}, we see that the spike trains change from trial to trial. Firing patterns become more irregular if either the number of neurons in the network or the amplitude of the injected noise increases, until they become close to Poisson. However, even in this limit, the amount of injected noise needed is relatively small compared to the synaptic inputs, and so the spiking irregularity does not simply reflect large amounts of external noise (see Appendix \ref{appendix:noise}). Strictly speaking, the input-output function for such a network is no longer deterministic, but should rather be described by a distribution $p(y|x)$, with a mean that is prescribed by a convex function from $x$ to $y$.

\subsection{Irregular firing amplifies noise in the decoder nullspace}

\begin{figure}[t!]
\centering
  \includegraphics[scale=1]{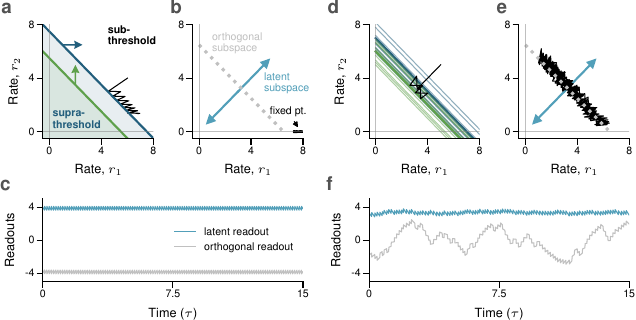}
  \caption{\label{firingrates}
   Irregular firing amplifies noise in the decoder nullspace in the inhibitory population.
    \textbf{a}: In the rank-1 inhibitory network, all neural thresholds are parallel in firing rate space (shown for two example neurons with rates $r_1$, dark blue, and $r_2$, green). For a deterministic network, only one of the neurons will fire (blue, $r_1$, dynamics in black), eventually reaching a fixed point (shown in panel \textbf{b}). \textbf{b}: The latent readout subspace (light blue) is orthogonal to the thresholds. The gray dotted line shows the subspace orthogonal to the readout (latent nullspace). \textbf{c}: Simulation of the steady-state spiking dynamics of the deterministic, two-neuron network from \textbf{a,b}. Both latent readout and activity projected into the orthogonal subspace are noise-free. \textbf{d}: For a network contaminated by noise, the neural thresholds fluctuate in firing rate space. As a consequence, both the green and blue neuron participate in the dynamics (black). \textbf{e}: With both neurons taking random turns in firing, the dynamics diffuse along the orthogonal subspace.
    \textbf{f}: For the network with noise, the readout is no longer deterministic, but the noise is relatively small. In comparison, activity projected into the orthogonal subspace fluctuates wildly.
    }
\end{figure}

The irregularity or noisiness of the individual spike trains of the network has important consequences for decoding. To visualize these consequences, we shift from the low-dimensional latent space back to the $N$-dimensional space spanned by the filtered spike trains or firing rates of the neurons, $\br = (r_1, r_2, \ldots, r_N)$.  We first observe that each neuron's threshold also describes a hyperplane in firing rate space. When the voltage reaches the threshold, the terms of the neuron's voltage equation (\Eq{NoisyThreshVoltage}) can be rewritten and solved for $y$ as
\begin{equation}
y = \sum_{j=1}^N D_j  r_j = \frac{T_i  - h*\eta(t) + \mu r_i(t)- F_i x}{E_i}, \label{eq:rate-hyperplane}
\end{equation}
which defines a hyperplane with normal vector $\bD=(D_1,\ldots, D_N)$ and offset as defined by the right-hand-side. The hyperplanes of different neurons have the same orientation, but different offsets. Two of these hyperplanes are shown in \Fig{firingrates}a for a two-neuron network. 

The dynamics of the network can similarly be understood in this space. The firing rate vector, $\br$, will leak towards zero in the absence of spiking (compare \Eq{filteredspikes}). When it hits one of the thresholds, the respective neuron, say neuron $k$, fires, and updates the $k$-th firing rate. In the absence of noise, the firing rate vector will always hit the same neuron (\Fig{firingrates}a), eventually reaching a fixed point (\Fig{firingrates}b,c). However, when the neurons' voltages are contaminated by noise, the thresholds randomly move around their default positions, and other neurons can eventually be the ones that fire (\Fig{firingrates}d). As a consequence, the firing rate vector starts to diffuse along the surfaces spanned by the neural thresholds (\Fig{firingrates}e). 

As a consequence of {\Eq{eq:rate-hyperplane}}, the latent readout $y$ is exactly orthogonal to this random diffusion along the threshold boundaries, and is therefore largely unaffected by it ({\Fig{firingrates}}b,e,f). However, if we project the neural activities onto a readout positioned in the orthogonal subspace, then we can see the large fluctuations from the random diffusion, which is caused by the random, irregular spiking or the neurons ({\Fig{firingrates}}f, gray curve). Accordingly, the recurrent connectivity of the network ensures that the latent readout remains relatively independent of the noise generated by irregular activity, which will only appear in orthogonal directions. This result is similar to previous findings \citep{boerlin2013predictive,landau2021macroscopic, depasquale2023centrality}, and may also relate to controllability of network dynamics {\citep{kao2019neuroscience}}.

\subsection{The rank-2 EI network is inhibition-stabilized and balanced}

\begin{figure}[t!]
\centering
  \includegraphics[scale=1]{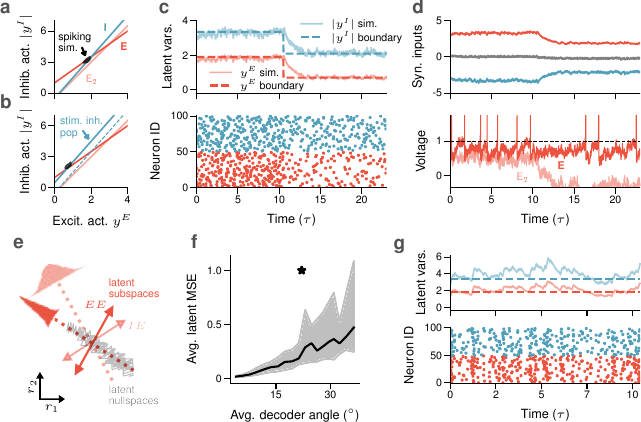}
  \caption{\label{EI-noise-mistuned}
  Irregular firing, balance, and noise control in the rank-2 EI network. \textbf{a}: Population boundaries for a homogeneous population of $N^E=50$ excitatory (E, red) and $N^I=50$ inhibitory (I, blue) neurons with fully aligned cross-connections and stable spiking dynamics at the boundary crossing (black); one additional inactive excitatory neuron with different parameters is shown ($E_2$, pink). \textbf{b}: When all inhibitory neurons are stimulated, the original boundary (dashed blue) moves upward (solid blue), and the fixed point is shifted down (spiking sim. in black). \textbf{c}: Latent readouts (top) and spike rasters (bottom) of the network over time; the inhibitory population is stimulated at $t=10$, which shifts the boundary intersection (dashed lines) and the spiking dynamics (solid faded lines). \textbf{d}: Balanced inputs (top) and voltage (bottom, red) for an example excitatory neuron from the homogeneous population (red); the additional excitatory neuron ($E_2$, pink) is hyperpolarized. \textbf{e}: Illustration of how mistuned decoders can amplify noise, shown in ($r_1^E,r_2^E$) firing rate space (compare with \Fig{firingrates}); spiking dynamics (faded gray) cause fluctuations in the nullspace of $D^{EE}$ (dotted red), which will be correlated with the latent readout $D^{IE}$ (solid pink). \textbf{f}: Average mean squared error (MSE) between true boundary and spiking simulation as a function of mistuning (angle between decoders; shading indicates standard deviation). \textbf{g}: Example simulation for misaligned decoders (indicated by star in panel \textbf{f}), resulting in noisy fluctuations.
    }
\end{figure}

We now return to the coupled excitatory-inhibitory network, where we have one latent variable for the excitatory and one for the inhibitory population, $y^E$ and $y^I$, respectively. We first consider a noisy version of the simple two-neuron setup from \Fig{EI_bounds}d, replacing each neuron by a population with $N^E=50$ and $N^I=50$ identical neurons, and adding a small amount of voltage noise (and self-reset) to each neuron, following \Eq{NoisyThreshVoltage}. The boundary crossing in latent space remains unchanged (\Fig{EI-noise-mistuned}a), but is composed of two homogeneous populations. We then add one additional excitatory neuron (labelled $E_2$, \Fig{EI-noise-mistuned}a) with a different boundary --- this neuron does not contribute to coding, but is illustrative of how boundaries geometrically relate to voltage, as we discuss below. Finally, to illustrate the inhibitory stabilization of the network, we consider the effect of a positive current stimulation to all neurons of the inhibitory population, a protocol that typically results in a ``paradoxical effect'' in inhibition-stabilized networks \citep{sadeh2021inhibitory}. Such a stimulation shifts the inhibitory boundary upwards, and should result in a new steady state of the latent variable activity in which both populations have reduced activity (\Fig{EI-noise-mistuned}b).

We simulate this larger, noisier network, applying the inhibitory perturbation for the second half of the simulation time (\Fig{EI-noise-mistuned}c). We observe stable coding of the two latent variable readouts as predicted by the boundary crossing, with asynchronous irregular spiking activity. Furthermore, the stimulation of the inhibitory population causes a decrease in the activity of both populations, confirming the network as inhibition-stabilized. We note again that this relies upon our assumption that the inhibitory population is faster than the excitatory population (see Section~\ref{ratesection} and Discussion). Next, by separating the positive and negative contributions to the voltage, we see that the synaptic inputs are roughly balanced, such that neurons in the network fire irregularly due to positive input fluctuations (\Fig{EI-noise-mistuned}c).

Lastly, we note that the additional excitatory neuron, $E_2$, does not fire any spikes --- its voltage remains hyperpolarized ({\Fig{EI-noise-mistuned}}d, bottom), with the relative amount of hyperpolarization being roughly proportional to the distance between its threshold and the current latent dynamics ({\Fig{EI-noise-mistuned}}a,b). We thus see that neurons in the network will generally be in one of two dynamic states --- they will either be contributing to coding by firing spikes, in which case their voltages will be balanced, or they will be silent with hyperpolarized voltages (see Discussion).

\subsection{Mistuned Cross-connections amplify noise due to irregular firing}

In this last sub-section, we now finally consider the general case of the EI network where each population utilizes independent decoders to read out the latent variables. That is, in contrast to the previous shared decoder case in \Eq{EI_WIE} and \Eq{EI_WEI}, we now define the cross-connections as 
\begin{align}
W^{IE}_{ij} &= E^{IE}_i D^{IE}_j\\
W^{EI}_{ij} &= E^{EI}_i D^{EI}_j, 
\end{align}
where $D^{IE}_j$ and $D^{EI}_j$ are potentially distinct from the self-decoders $D^{E}_j$ and $D^{I}_j$, respectively. With this redefinition of \Eq{EI_WIE} and \Eq{EI_WEI}, we are now looking at a rank-4 EI network, in which all cross- and self-connections can be arbitrary (sign-constrained) rank-1 matrices. The cross-connection matrices can now be ``misaligned'', and so they will generally read out a mixture between each population's own latent readouts, $y^E$ and $y^I$, and the orthogonal subspace discussed above (\Fig{EI-noise-mistuned}e). To formalize these cross-connection readouts, we introduce two new latent variables,
\begin{align}
y^{IE} &= \sum_{j=1}^{N^E} D^{IE}_j r^E_j\\ 
y^{EI} &= \sum_{j=1}^{N^I} D^{EI}_j r^I_j. 
\end{align}
These misaligned readouts can be explicitly split into a part that aligns with the self-decoders, and a part that captures noise from the null space. For example, let us focus on the excitatory readout of the inhibitory population. Writing $\bD^X=(D^X_1,\ldots,D^X_N)$ for the vector of decoders for all four cases, $X\in\{E,EI,IE,I\}$, and assuming for simplicity that the four decoding vectors are normalized, $(\bD^X)^\top\bD^X=1$, we obtain 
\begin{align}
y^{IE} &= (\bD^{IE})^\top\br^E\\
& =(\bD^{IE})^\top
\Big( \bD^E(\bD^E)^\top + \bI-\bD^E(\bD^E)^\top\Big)
\br^E\\
& = ({\bD^{IE}})^\top\bD^E\, y^E\, +\, {\bD^{IE}}^\top\big(\bI-\bD^E(\bD^E)^\top\big)\br^E\\
& = \alpha y^E + (\bD^{IE}-\alpha\bD^E)^\top\br^E.
\end{align}
Here, the first term captures a decreased readout of the correct excitatory latent, $y^E$, since $\alpha=(\bD^{IE})^\top{\bD^{E}}\leq 1$, and the equality sign only applies when $\bD^{IE}=\bD^E$. In turn, the second term captures the random fluctuations from the orthogonal subspace. As the decoders $\bD^{IE}$ and $\bD^E$ become less and less aligned, $\alpha$ decreases and the relative power of signal and noise shifts towards noise (\Fig{EI-noise-mistuned}f,g; see also \cite{landau2021macroscopic}). The respective networks will thereby amplify noise and contaminate the signal. When the two readouts become orthogonal, $\alpha=0$, the inhibitory population receives only noise and no longer any signal, $y^E$. In this extreme case, the stability will be compromised (not shown).

%%%%%%%%%%%%%%%%%%%%%%%%%%%%%%%%%%%%%%%%%%%%%%%%%%%%%
%%%%%%%%%%%%%%%%%%%%%%%%%%%%%%%%%%%%%%%%%%%%%%%%%%%%%
\section{Rate networks can approximate spiking networks in the latent space}\label{ratesection}

Up to now we have studied an idealized spiking network model that ignores several basic properties of real neurons. Most notably, we have assumed that synaptic input currents are instantaneous, i.e., they arrive immediately and are infinitely short, and we have imposed that inhibitory neurons fire before excitatory neurons. As we will see, relaxing these assumptions will influence how sharp the boundary is and how accurately the input is mapped onto the output. In doing so, we arrive at a relationship between spiking dynamics and the smoothed, trial-averaged firing-rate dynamics that are typically studied in rate networks.

\subsection{Slower synapses generate a finite boundary}

\begin{figure}[t!]
\centering
  \includegraphics[scale=1]{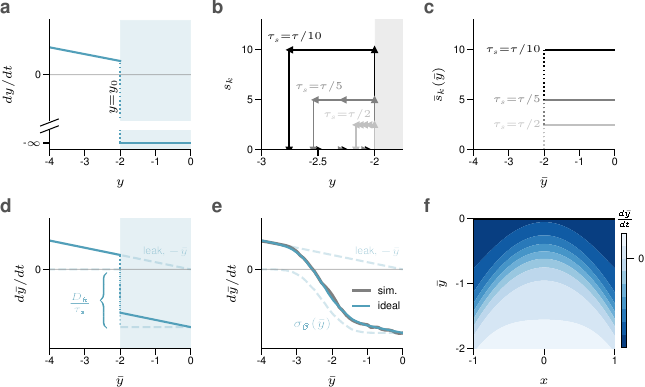}
  \caption{\label{Iysoft}
  From hard to soft boundaries in the inhibitory population.
    \textbf{a}: Plot of latent $y$ versus its derivative $dy/dt$ for a synaptic delta-current (analogous to a vertical slice through the inhibitory boundary from \Fig{Ibounds1}a,c). The threshold of the neuron sitting at $y=y_0=-2$ causes an infinitely steep boundary, and divides the subthreshold (white) from suprathreshold (blue) areas. 
    \textbf{b}: Introduction of finite-time, square pulse synaptic dynamics, with time constant {$\tau_s$} (shown for three values of {$\tau_s$} in black, gray, and silver), results in oscillatory dynamics in ($y$, $s_k$)-space, with jumps in $s_k$ of size $1/\tau_s$. Arrows indicate equally-spaced time intervals.
    \textbf{c}: The time-averaged dynamics of $y$, denoted $\bar{y}$, can be qualitatively approximated by assuming fixed synaptic input of amplitude $1/\tau_s$ when the boundary is crossed (at $\bar{y}=\bar{y}_0$), resulting in $s_k(\bar{y})$ as a scaled Heaviside function.
    \textbf{d}: The resulting plot of the average latent $\bar{y}$ versus its derivative for the approximation to the square pulse synapse, $s_k(\bar{y})$, from (c), where the jump in the derivative is a negative Heaviside function scaled by $D_k/\tau_s$.
    \textbf{e}: Averaging the spiking dynamics for the threshold boundary from (d) jittering with white noise (gray, sim.) results in a soft boundary well-approximated by a sigmoid plus leak (blue). Trials were simulated with random initial conditions for $y$ in the interval $(-4,4)$. Comparison with an `ideal' sigmoid function (blue), simulated as the average over the hard boundaries in (d) with Gaussian jitter.
    \textbf{f}: The soft boundary equivalent to the hard inhibitory network boundary from \Fig{Ibounds1}b-d and \Fig{simulnoise}, in which neurons are arranged along a quadratic function.
    }
\end{figure}

First, we will examine how the boundary is affected when we change our assumptions about postsynaptic currents. For simplicity, we will first return to a single, rank-1 inhibitory population. Previously, we demonstrated that the dynamics of $y$ are characterized by the combination of a leak and an infinitely steep boundary (compare \Eq{yIneurons}, but assuming a one-dimensional input),
\begin{equation} 
\dot{y} = -y + \sum_{i=1}^N D_i \, I\big( F_ix + E_iy - T_i\big).\label{Iyneuronsrewrite}
\end{equation}
For a fixed input $x$, we illustrate this infinitely steep boundary in \Fig{Iysoft}a, where we plot $y$ versus $dy/dt$ (consider this as a vertical slice through \Fig{Ibounds1}a-c). The dynamics at the boundary are oscillatory, with leak dynamics, followed by instantaneous spiking (see \Fig{Ibounds1}d). The latent readout ({\Eq{readoutdydt}}) can also be expressed in terms of the respective spike times, $t_i^f$,
\begin{equation}
\dot{y}=-y+\sum_{i=1}^N\sum_f D_i \delta(t-t_i^f),\label{y_delta}
\end{equation}
where $t_i^f$ is the $f$-th spike fired by the $i$-th neuron. Here, the delta-function models the effect of a spike on the latent variable. Biophysically, the latent variable $y$ is linearly related to the voltage (see \Eq{IV}) and so the delta-function corresponds to an infinitely short, postsynaptic current.

In reality, of course, the postsynaptic current generated by a spike will last for a finite amount of time. We can explicitly include these synaptic dynamics by replacing the delta function in \Eq{y_delta} with a model of finite synaptic dynamics, such as exponential decay or a square pulse \citep{gerstner2014neuronal}. Taking the square pulse as an example, we write its synaptic dynamics as
\begin{equation}
\alpha(t) = \frac{1}{\tau_s}H(t)H(\tau_s-t),\label{square-pulse}
\end{equation}
where the synaptic time constant $\tau_s$ is given in units of the membrane time constant, since $\tau=1$ in our case. \Eq{square-pulse} scales $\alpha(t)$ with $1/\tau_s$, such that it always integrates to one, and recovers the Dirac delta in the limit as $\tau_s\rightarrow0$. Repeating the derivations in Sections 2.1-2.2, the latent dynamics becomes (compare \Eq{readoutdydt})
\begin{align}
  \dot{y}&=-y + \sum_{i=1}^N\sum_f D_i \alpha(t-t_i^f) \\
         &=-y + \sum_{i=1}^N D_i s_i(t)
\end{align}
where in the last step we re-defined the `spike train,' $s_i(t)$, which should now primarily be interpreted as a synaptic current input, consisting of a sum of postsynaptic currents.

We now consider a section of the boundary formed by one neuron, say neuron $k$, and visualize the dynamics in $(y,\,s_k)$-space for the square pulse synapse model (\Fig{Iysoft}b; shown for three values of $\tau_s$, black, gray, and silver). The system relaxes into a steady-state oscillation at the boundary. However, the presence of the synaptic current, $s_k$, complicates the dynamical picture by adding another dimension, so that each neuron  now depends on two dynamical variables.

To simplify the dynamics, we will aim to qualitatively describe an approximation of $y$, which we will denote as $\bar{y}$, by replacing the temporal dynamics in $s_k(t)$ with a simple static function $\bar{s}_k(\bar{y})$. We will take advantage of the fact that the square pulse synapse model only takes on one of two discrete values. Qualitatively, we see that $s_k$ sits at zero until the boundary is reached, when it jumps up to a non-zero value momentarily, $s_k\rightarrow 1/\tau_s$, at which point $y$ moves negatively back into the subthreshold area, until $s_k$ jumps back to zero. The time scale $\tau_s$ affects the size of the jump in $s_k$, as well as the size of the change in $y$. This change in $y$ is bounded by $D_k$ (if $\tau_s\to 0$), but gets smaller and slower as $\tau_s$ increases. In order to replace this oscillatory dynamics by an effective time-averaged dynamics, we assume a functional form of $\bar{s}_k$ as a scaled Heaviside function --- it jumps up by $1/\tau_s$ whenever the boundary is crossed, and jumps down to zero when $y$ moves back into the subthreshold set (\Fig{Iysoft}c; three values of $\tau_s$ shown, corresponding to \Fig{Iysoft}b). As a consequence, we can approximate each neuron's $s_i$ as
\begin{equation}
\bar{s}_i(\bar{y}) = \begin{cases}
    1/\tau_s & \text{if }F_ix + E_i\bar{y} - T_i > 0 \\
    0 & \text{otherwise}.
\end{cases}
\end{equation}
This approximation allows us to effectively remove the history dependence of the synapses. Back at the level of the latent variable dynamics, we can now write 
\begin{align}
\dot{\bar{y}} &= -\bar{y} + \sum_{i=1}^N D_i\bar{s}_i(\bar{y})\nonumber\\
&= -\bar{y} + \sum_{i=1}^N \frac{D_i}{\tau_s} H(F_ix + E_i\bar{y} - T_i).
\end{align}
We have thereby replaced the previously oscillatory dynamics at the boundary with a single fixed point, given when the boundary is reached (strictly speaking, the fixed point of the average dynamics does depend slightly on the time scale $\tau_s$, and will be different from the original $y_0$, but we have here ignored this effect for simplicity).

As illustrated in \Fig{Iysoft}d, the infinitely steep boundary  has become a finite-sized boundary. We can also see that the size of this boundary grows with the size of the decoder, $D_k$, and is inversely related to the timescale of the synaptic dynamics, $\tau_s$. Moreover, when we add a second, identical neuron to the picture, both neurons will fire two spikes per oscillation cycle, so that the effective size of the boundary doubles (not shown). More generally, the size of the boundary therefore scales linearly with the number of neurons at the boundary, as long as their thresholds are sufficiently closely spaced.

\subsection{Noise causes a soft boundary}

Let us next see how the boundary is affected when we add noise to the system. 
In Section~\ref{noisecontrol}, we saw that the main effect of voltage or current
noise was to randomly move the thresholds of the neurons. Similar effects
can be obtained when synaptic weights or voltage resets slightly deviate from their idealized set points \citep{calaim2022geometry}. 

In \Fig{Iysoft}e, we simulated a neuron with normally-distributed threshold noise added and we plot the trial-averaged output against the trial-averaged derivative. With a slight abuse of notation, we will refer to this trial-averaged latent output as $\bar{y}$ as well. We see that the main effect of the noise is to soften the boundary. Mathematically, the reason for this softening is that a Heaviside function with input subject to Gaussian noise takes the shape of the sigmoidal error function (erf). Due to its common use in rate networks and qualitative similarity to the error function, we choose to use the logistic function,
\begin{equation}
\sigma_\beta(u) = \frac{1}{1+e^{-\beta u}},\label{logistic}
\end{equation}
where the parameter $\beta$ determines the steepness and is inversely proportional to the noise. For $\beta\rightarrow \infty$, we recover the deterministic Heaviside function. Apart from the parameters mentioned before---strength of the decoding weights, the inverse of the synaptic time scale, number of neurons around the boundary---the steepness of the boundary therefore also depends inversely on the level of the noise. When we now put this picture back into the full input-output, ($x$-$y$)-space as before, we can visualize the same network boundary as in \Fig{Ibounds1}b-d and \Fig{simulnoise}, but now for the soft case, which is shown in \Fig{Iysoft}f.

We thus have the following dynamics for the trial-averaged activity of a network with finite synaptic dynamics,
\begin{equation}
    \dot{\bar{y}} = -\bar{y} + \sum_{i=1}^N \frac{D_i}{\tau_s}\sigma_\beta(F_ix + E_i\bar{y} - T_i).\label{trialavy}
\end{equation}
This equation therefore describe the boundary by a single stable fixed point with locally asymmetric dynamics: fast attraction due to a steep boundary on one side and shallow attraction due to the leak on the other side (\Fig{Iysoft}e). The slower the synapses and the larger the noise, the less strong this asymmetry becomes, until the notion of a boundary ceases to be useful in describing the system dynamics.

\begin{figure}[t!]
\centering
  \includegraphics[scale=1]{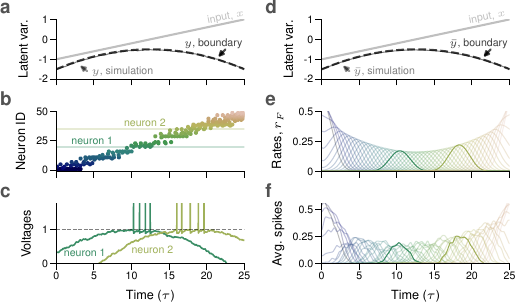}
  \caption{\label{Iysoft-examples}
    The inhibitory network with slow synapses. \textbf{a}-\textbf{c}: The network of $N=50$ neurons from \Fig{simulnoise} with slow synapses (square pulse; $\tau_s=0.5\tau$) yields stable coding (\textbf{a}) and irregular firing (\textbf{b,c}) (refer to {\Fig{simulnoise}} caption for details). \textbf{d}-\textbf{f}: A rate version of the network also has stable coding (\textbf{d}) and has single-unit rate activities (\textbf{e}) that reproduce the trial-averaged activity of the spiking network (\textbf{f}; Gaussian- smoothed spikes for $100$ trials of the system in (\textbf{a}-\textbf{c})).
    }
\end{figure}

To illustrate this softer boundary at the population level, we simulated the network from \Fig{Iysoft}f with the same setup from \Fig{simulnoise}, but now with finite square-pulse synaptic dynamics ($\tau_s=0.5\tau$, see \Fig{Iysoft-examples}a-c). We see that in this case, the latent readout still closely follows the threshold boundary (\Fig{Iysoft-examples}a), and the irregular firing and trial-to-trial variability are still retained (\Fig{Iysoft-examples}b,c). This demonstrates that the finite, soft boundary can still be geometrically visualized in the same way as the idealistic, infinite boundary from before, and that fragility issues previously associated with spike-coding networks \citep{chalk2016neural, rullan2020poisson, calaim2022geometry} can be alleviated in this regime.

\subsection{The latent variables of spiking and rate networks are equivalent}

The dynamics of the trial-averaged latents has a simple relation to classical firing rate networks. Let us use the notation $r_{\text{F},i}$ to denote the activation of a unit $i$ of a firing-rate network. Assuming a one-dimensional input $x$, the  equation for an all-inhibitory rate network can be written as
\begin{equation}
\dot{r}_{\text{F},i}= -r_{\text{F},i} + \sigma_\beta\big( F_i x + \sum_j W_{ij} r_{\text{F},j} - T_i \big),
\end{equation}
where $\sigma_\beta(\cdot)$ is the sigmoidal activation function of each neuron, \Eq{logistic}, and the threshold term $T_i$ simply serves as a constant negative offset. Setting the inhibitory connection matrix to be rank-1, $W_{ij}=E_iD_j$, with $D_j\leq 0$, and defining the readout as $y_{\text{F}}=\frac{1}{\tau_s}\sum_i D_i r_{\text{F},i}$, we can compute the derivative of $y_\text{F}$ to obtain, 
\begin{equation}
\dot{y}_\text{F}=-y_\text{F} +\sum_{i=1}^N \frac{D_i}{\tau_s}\sigma_\beta\big(
F_i x + E_i y_\text{F}- T_i \big). \label{rateeq}
\end{equation}
We note that this equation is identical to \Eqs{yavI}{yaVE}, if we identify the readout $y_F$ with the trial-averaged readout $\bar{y}$ from the spiking network. Indeed, when we simulate this firing-rate version of the network, we see that it tracks the boundary well, and the firing rates of the individual neurons closely match the trial-averaged rates of the spiking network (\Fig{Iysoft-examples}d-f)).

Formally, we have therefore shown that the dynamics of the latent variables of a rank-1 (inhibitory) firing rate network are equivalent to the trial-averaged dynamics of a rank-1 (inhibitory) spiking network, as long as parameters are set such that the networks generate a clear boundary (see section 4.6 and Discussion).

\subsection{The E-I boundaries become nullclines in the firing-rate limit}

\begin{figure}[t!]
\centering
      \includegraphics[scale=1]{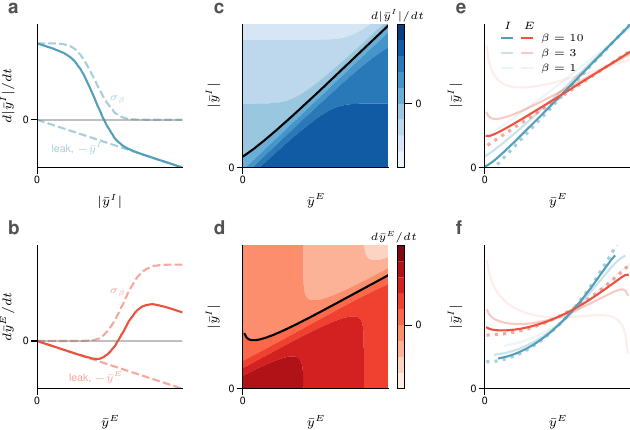}
      \caption{\label{EIdynamics}
      Dynamics of the rank-2 EI network with soft boundaries.
        \textbf{a}: Stable inhibitory dynamics (solid line), composed of a sigmoidal boundary and a leak (dashed lines). Replotted from \Fig{Iysoft}e, using the absolute value of the inhibitory output, $|\bar{y}^I|$.
        \textbf{b}: Unstable excitatory dynamics, again composed of a sigmoidal boundary and a leak.
        \textbf{c}-\textbf{d}: The soft inhibitory and excitatory boundaries for the rank-2 EI network in 2d latent space (homogeneous populations). Nullclines are indicated by the black curves.
        \textbf{e}: I and E nullclines of the homogeneous rate-based rank-2 EI system for three different values of $\beta$, controlling the slope of the sigmoids. The boundaries of the spike-based system are shown as dotted lines for comparison.
        \textbf{f}: I and E nullclines for a \emph{heterogeneous} rank-2 EI system, in which neurons are arranged along quadratic functions as in \Fig{EI_bounds}c,f.
  }
  \end{figure}

We now return to the rank-2 E-I network, and consider the effect of slower synapses here. First, we adopt the arguments made above for the stable inhibitory boundary and apply them to the unstable excitatory boundary. We then see that the two boundaries differ by a simple sign flip in the $(|\bar{y}|,\dot{\bar{y}})$-plot (compare ({\Fig{EIdynamics}}a against {\Fig{EIdynamics}}b). Note that the output of the inhibitory population is negative in our convention (compare \Fig{Iysoft}c), and we here plot its absolute value to allow better comparison with classical plots of E-I networks.

When combining the two populations, we once more assume that they share the same decoders, following \Eq{EI_WII}-\Eq{EI_WEE}. Starting from \Eq{rateeq}, we then obtain the following equations for the trial-averaged dynamics,
\begin{align}
 \dot{\bar{y}}^I  &= -\bar{y}^I +\sum_{i=1}^{N^I} \frac{D^I_i}{\tau^I_s} \sigma_\beta \big( F_i^I x + E_i^{IE}\bar{y}^E + E_i^{II}\bar{y}^I  -T_i^I \big) \label{yavI}\\
 \dot{\bar{y}}^E &= -\bar{y}^E + \sum_{i=1}^{N^E} \frac{D_i^E}{\tau^E_s} \sigma_\beta\big( F_i^E x + E^{EE}_i\bar{y}^E + E^{EI}_i\bar{y}^I -T_i^E \big).\label{yaVE}
\end{align}
Compared to the original equation for the spiking network, \Eqs{Idyn_EI}{Edyn_EI}, we again see that we have simply replaced the indicator function $I(\cdot)$ by a sigmoidal function, $\sigma_\beta(\cdot)$, divided by the time scales of the synaptic current pulses, $\tau^I_s$ and $\tau^E_s$, respectively.

We illustrate these soft boundaries in the two-dimensional latent  space, $(|\bar{y}^I|,\bar{y}^E)$,  for two neurons (or two homogeneous populations) in {\Fig{EIdynamics}}c,d. At a height of $(|d\bar{y}^I/dt|,d\bar{y}^E/dt)=0$, we retrieve the nullclines of the dynamics, as illustrated by black lines in {\Fig{EIdynamics}}c,d. We observe that the exact shapes and positions of the nullclines depend on the slope parameter $\beta$ of the sigmoids, which approach the spiking boundaries (\Fig{EIdynamics}e, dotted lines) for large $\beta$. 

For heterogeneous networks, we previously approximated each population's boundary as lying along a smooth convex curve, similar to {\Fig{EI_bounds}}f. When we soften the boundary, its overall shape is both determined by the individual neural thresholds, as well as the overall smoothness of the sigmoidal boundary (\Fig{EIdynamics}f).

\begin{figure}[t!]
\centering
      \includegraphics[scale=1]{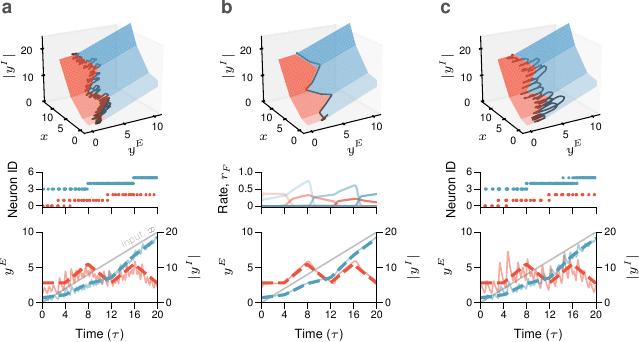}
      \caption{\label{slowEI-UFA}
        Function approximation in the rank-2 EI network with slow (spike-based) synapses and rate dynamics (compare with and refer to {\Fig{EI-function-approx}} caption). \textbf{a}: The example from \Fig{EI-function-approx} with slow synapses ($\tau^E_s=\tau/4$, $\tau^I_s=\tau/10$) yields a reliable, yet noisier approximation to the boundary crossing.
        \textbf{b}: A rate-based version of the network with stable, reliable dynamics, and with rates that match the trial-averaged activity of the spiking network (not shown).
        \textbf{c}: Slowing down the inhibitory synapses sufficiently ($\tau^E_s=\tau/4$, $\tau^I_s=\tau/4$) leads to more oscillatory dynamics, with less reliable coding.
  }
  \end{figure}

We illustrate the functionality of the rank-2 EI network with finite synapses by repeating the function approximation example from \Fig{EI-function-approx}b-d. As illustrated in \Fig{slowEI-UFA}a, function approximation is made noisier with slowed-down synapses, but it is still reliable. While we no longer apply the ad hoc rule that inhibitory neurons fire first, the inhibitory synaptic dynamics should ideally be faster than the excitatory dynamics, in order to ensure stability (in this case $\tau_E=\tau/4$ and $\tau_I=\tau/10$; see Discussion).

Just as in the previous section, we can similarly derive the above equations by starting with a rank-2 firing rate network. When simulating such a network, we see that the latent readouts smoothly follow the intersection of the two boundaries, and rates will approximate the trial-averaged activities of the spiking network (\Fig{slowEI-UFA}b).

\subsection{Synaptic time scales impact the accuracy of the output}

Beyond making the boundary softer, the combination of mistuned network parameters and realistic synaptic currents can also influence the spiking dynamics and thereby the accuracy of the network's input-output mapping. Recall that the output in the original network fluctuates between the threshold boundary and the reset boundary (see \Fig{Ibounds1}c). However, when synaptic currents are not instantaneous, and when the thresholds are not fixed, then the output can sometimes cross multiple thresholds, and multiple neurons fire.  In this case, the output $y$ does not simply bounce back to the reset boundary, but moves further into the subthreshold regime. This will be especially true if the synaptic time constants are very slow, or if there are significant delays in axonal conductance. These resulting oscillatory effects in $y$  decrease the accuracy with which the input is mapped onto the output. An example is shown in \Fig{slowEI-UFA}c, in which $\tau^E_s=\tau^I_s=\tau/4$. Such deviations will also disrupt the efficiency of the code in terms of the number of spikes needed to represent a particular output level \citep{deneve2016efficient} (see Discussion).

From a computational perspective, this loss of accuracy and efficiency is not desirable. These problems will likely be mitigated in higher-rank networks \citep{calaim2022geometry}, and thus may be less severe in scaled-up systems (see Discussion). From a biological perspective, simultaneous crossing of multiple thresholds causes synchronous firing, which may relate to oscillatory dynamics in the brain \citep{chalk2016neural}, suggesting that some residual effects may be present  in real circuits.

\subsection{The spike-rate equivalence requires a latent boundary}

We note that the above derivations are restricted to networks that have a block-wise rank-1 structure, i.e., in which both self-connectivity and cross-connectivity of the excitatory and inhibitory subnetworks is rank-1. However, not all networks with this structure will permit a transition from spiking to rate networks as described here.

First, we have assumed that the cross-connections have the same decoders as the self-connections of each population (compare \Eqs{EI_WII}{EI_WEE}). As explained in Section~3, when the cross-connections are not aligned, the spiking network will accumulate noise, and a match between trial-averaged spiking networks and rate networks can no longer be guaranteed. This restriction on cross-connection alignment is usually not followed in the design of firing rate networks.

Second,  parameters in classical firing rate networks are usually chosen such that individual neurons operate in the full regime of their input function, from below threshold to large firing rates and even into saturation. The underlying reason is always that the neural activation functions serve as basis of an underlying function approximation scheme (compare \Fig{puzzle-fig}a--c). Apart from feedforward networks, examples include line attractor networks in which neural thresholds are staggered in latent space so that the addition of all neural activation functions  precisely counters the leak  \citep{seung1996brain,eliasmith2003neural}, or decision-making models that use saturation to obtain bistability \citep{mastrogiuseppe2018linking}. 

In contrast, in the spiking networks here, neural thresholds are aligned in latent space such that they join to form a boundary. As a consequence, each neuron touches the boundary at some point, and its activation function enforces the boundary. Indeed,     parameters need to be chosen such that the boundary is so steep as to be basically unsurmountable by the latent variable. The boundary becomes particularly steep when the synaptic currents, $\tau_s$, become sufficiently short. Apart from that constraint, the precise shape of the boundary function does not matter.  

Importantly, this operating regime is crucial in order to 
obtain balanced synaptic current inputs into the neurons (compare \Fig{puzzle-fig}d,e). In the rate-domain, our perspective bears a resemblance to other rate-based models featuring non-saturating nonlinearities \citep{miller2002neural, hansel2002noise, ahmadian2013analysis, rubin2015stabilized, hennequin2018dynamical}.
Indeed, we would reinterpret the respective neural activation functions as implementing soft boundaries.

%%%%%%%%%%%%%%%%%%%%%%%%%%%%%%%%%%%%%%%%%%%%%%%%%%%%%
%%%%%%%%%%%%%%%%%%%%%%%%%%%%%%%%%%%%%%%%%%%%%%%%%%%%%
\section{Discussion}

In this article, we have proposed a new framework for constructing and interpreting a broad class of spiking neural network models based on spike-threshold boundaries. These boundaries come from the inherent inequality at the core of each neuron --- the voltage threshold --- which triggers a spike. Classic approaches usually seek to eliminate the threshold inequality by, e.g., filtering or averaging spike trains, and to replace it with a continuous activation function \citep{dayan2005theoretical, gerstner2014neuronal}. We take a different approach here. By defining a one-dimensional latent variable readout, we show that the spike-threshold boundaries of rank-1 networks can be visualized in input-output space, with each neuron's threshold forming an affine boundary. A population of neurons delineates a convex boundary separating subthreshold from suprathreshold areas, and for an inhibitory population this boundary creates a stable input-output function. For an excitatory population, the boundary is instead unstable, but can then be stabilized in a coupled EI network, where the dynamics sit at the crossing of the two boundaries. The boundary between sub- and supra-threshold regimes thus becomes the center of spike-based computation, and distinct from the ``feature detector'' perspective (\Fig{puzzle-fig}).

The mathematics of systems with inequalities and hard boundaries has a rich history in constrained convex and non-convex optimization \citep{boyd2004convex}. Previous work on spike coding has demonstrated how the dynamics of spiking networks can be mapped onto convex optimization problems such as quadratic programming \citep{barrett2013firing, mancoo2020understanding}, yielding stable, convex input-output transformations. We extended this work here by showing that coupled EI networks are capable of non-convex function approximation, with links to difference of convex (DC) programming \citep{horst1999dc, lipp2016variations}. Similarly, previous work has linked EI networks to minimax optimization \citep{seung1997minimax, li2020minimax}, but with a focus more on attractor dynamics rather than function approximation. Overall, the analogy between spiking thresholds and constraint boundaries offers a fruitful perspective that may lead to new insights and algorithms that harness the power of convex optimization \citep{boyd2004convex}. Moreover, such techniques have recently caught on in deep learning in the form of deep implicit layers, offering potential avenues for learning and scaling up our framework \citep{amos2017optnet, gould2021deep}.

In the work presented here we limited ourselves to small populations of neurons with rank-1 connectivity. Though increasing the number of neurons in rank-1 (or rank-2 EI) networks may enable closer and closer approximations to arbitrary continuous 1-d functions, 
the $N\rightarrow\infty$ limit of rank-1 networks is likely too limiting to relate to biological spiking networks. Furthermore, though we do not demonstrate it here, this larger-scale limit requires the absence of noise and communication delays, in addition to having very large synaptic input currents. Instead, to stay within a biologically realistic and workable regime, scaling up the number of neurons should coincide with scaling up the rank of the connectivity \citep{calaim2022geometry}. 
Fortunately, many of the intuitions about spike thresholds and convex boundaries transfer to the higher-dimensional input-output spaces of arbitrary rank-K networks. Furthermore, many signatures of biological activity regimes discussed here, such as irregular firing and balance, become more robust in this higher-dimensional regime \citep{calaim2022geometry}.

The rank-1 and rank-2 spiking networks studied here can be seen as a particular generalization of spike-coding networks (SCNs) \citep{boerlin2013predictive, deneve2016efficient}, with the spike-threshold boundary being a generalization of the concept of a bounding box put forward in \citet{calaim2022geometry}. As we demonstrate, this generalization frees SCNs from the constraint of linear computation without evoking dendritic or synaptic nonlinearities \citep{thalmeier2016learning, alemi2018learning, nardin2021nonlinear, mikulasch2021local}. We can reinterpret the bounding box as being composed of two components --- local regions of the boundary are stable, with lateral inhibition between neurons, but globally the closed nature of the box may lead to errant positive feedback and thus fragility in the form of epileptic (``ping-pong'') behavior \citep{chalk2016neural, rullan2020poisson, calaim2022geometry}. Our framework eliminates this previous limitation, fully separating the positive and negative feedback loops in the network. Despite these differences, many of the desirable biological properties and predictions of SCNs are preserved here, including robustness to cell death and inhibition, firing irregularity, E/I balance, and coding efficiency. While it is out of the scope of this current study to explore these properties in detail, in future work it will be interesting to determine how the EI nature of this model leads to unique biological predictions, especially in higher-dimensional networks. More generally, the spike-coding framework has been extended in various ways for other computational aims (e.g., \cite{slijkhuis2022closed, masset2022natural}), as well as to incorporate additional biologically-plausible details or activity regimes (e.g., \cite{schwemmer2015constructing, koren2022biologically, safavi2023signatures}), and it will be interesting to see how our perspective here may be integrated with these other studies. Taking into account slower synapses with delays, we are able to relate our work to the dynamics of rate networks, similar to previous approaches \citep{kadmon2020predictive, rullan2020poisson}. However, our aim here is not to emulate a rate-based system with spikes, but rather to demonstrate the relationship between hard spike boundaries and soft rate boundaries.

One major biological implication of our work is the natural distinction between excitatory and inhibitory populations. Many previous models, in contrast, typically start with Dale's law as a biological constraint, then attempt to achieve performance on par with an unconstrained network, and finally reflect on the potential benefits of such a constraint for, e.g., stability or robustness \citep{hennequin2014optimal, song2016training, brendel2020learning, cornford2020learning, haber2022computational}. Our framework instead puts forth a particular computational role for Dale's law --- it is the combination of positive feedback from the unstable excitatory boundary with negative feedback from the stable inhibitory boundary that enables flexible input-output mappings. This idea is reminiscent of so-called mixed-feedback systems in control theory, which are beginning to be explored in the neuroscience domain \citep{sepulchre2019control, sepulchre2022spiking}. In addition to the unique computational role of Dale's law in our model, the separation into excitatory and inhibitory populations also links the dynamical behavior of our model to previous work on EI networks, including inhibition-stabilized networks \citep{sadeh2021inhibitory}. 

A strict requirement of the rank-2 EI networks that we study is that inhibition should be faster than excitation in order to keep the dynamics tightly around the intersection of the two boundaries. While some evidence suggests that the synaptic time constant of inhibition may be slower than that of excitation (e.g., see discussion in \cite{fourcaud2002dynamics}), other factors, such as the localization of inhibitory synapses close to the peri-somatic zone \citep{freund2007perisomatic}, or cortical connectivity patterns (short-range for inhibition, long-range for excitation) suggest that inhibition could also be faster than excitation. In addition, self-stabilizing mechanisms within the excitatory population, such as spike-frequency adaptation, could make the speed of inhibition less important.

Overall, our work can be seen as a bridge between mechanistic and functional network models of the brain. Mechanistic models, perhaps best represented by balanced spiking networks \citep{van1996chaos,amit1997model,brunel2000dynamics}, capture many important features of cortical dynamics, including asynchronous irregular activity, excitation-inhibition balance, and correlated neural variability \citep{vogels2005signal, renart2010asynchronous, rosenbaum2017spatial, huang2019circuit}. However, such models, and especially their theoretical analysis, have typically been 
limited to linear computations in large  networks (infinitely many neurons in the mean-field limit, or 1000-10000 in practice). In contrast, our work focuses on non-linear computations in much smaller spiking networks (5-50 neurons).

While various frameworks have been proposed to obtain non-linear computations in the mean-field limit \citep{renart2007mean, roudi2007balanced, hansel2013short, lajoie2016encoding, ingrosso2019training, kim2021training}, one recent study parallels our work in several interesting ways. \cite{baker2020nonlinear} utilize networks with ``semi-balance'' or ``approximate balance'' in order to achieve nonlinear computation. Notably, they observe that active sub-populations retain balanced input, whereas others are hyperpolarized. Despite the size differences, this is precisely what our networks predict as well (\Fig{EI-noise-mistuned}d), and future work should explore links between this work and our proposed framework. More generally, this supports the notion that the average balance must be broken or loose enough to enable nonlinear computations \citep{ahmadian2021dynamical}.

Functional models of static input-output mappings are perhaps best represented by feedforward rate networks \citep{hunsberger2015spiking, yamins2016using}. However, such networks do not agree with the heavily recurrent connections found in the brain, they do not obey Dale's law, and they rely on neurons being feature detectors, which conflicts with the idea of balanced inputs (\Fig{puzzle-fig}). Here we have shown that function approximation can also be achieved in (low-rank) recurrent networks obeying Dale's law, and with balanced inputs. Typically, the study of low-rank networks has emphasized the generation of internal dynamics for decision making, memory, and motor control \citep{eliasmith2005unified, mastrogiuseppe2018linking, dubreuil2022role}. Though we have deliberately ignored dynamics here, some simple dynamical motifs such as bistability can be achieved even in the rank-2 EI network provided that there are multiple boundary crossings (not explored here). Higher-dimensional systems can have much richer dynamics in the latent space, as is already known from work on rate networks, and will be explored in future work. One aspect of dynamics that we do consider in this work is the existence of noise-amplifying or noise-suppressing connectivities. These properties have also been described in other low-rank networks \citep{landau2018coherent, landau2021macroscopic,depasquale2023centrality}.

Advances in learning algorithms have led to an explosion of research on training functional spiking network models \citep{abbott2016building, neftci2019surrogate}. Interestingly, many algorithms smooth over the spike-threshold nonlinearity using ``surrogate'' or pseudo-gradients \citep{bellec2018long, neftci2019surrogate, zenke2021remarkable}, or directly train on rate dynamics \citep{depasquale2023centrality}, which may smooth out or prevent the development of well-formed latent boundaries. It will be interesting to develop scalable training algorithms for the networks discussed here, and to see how additional constraints during training may affect the development of spike-threshold boundaries.

In summary, this article puts forth a new perspective on spike-based computation. It offers the potential to better understand the computational regimes and benefits of the balanced state as observed in cortex, and presents a theoretical framework to construct functional spiking neural networks based on these insights. Finally, while the potential to scale up the framework to higher-rank networks with dynamical computations is still to be demonstrated, it may serve as a promising direction for scalable computation with spikes.

\section*{Acknowledgements}
\addcontentsline{toc}{section}{Acknowledgements}

We thank Alfonso Renart, Francesca Mastrogiuseppe, and  Sander Keemink for detailed comments on the manuscript. We thank all members of the Machens Lab for helpful discussions and feedback, especially Allan Mancoo for  discussions on convex optimization, and Bertrand Lacoste for discussions pertaining to decomposition into the difference of two convex functions. This work was supported by the Funda\c c\~ao para a Ci\^encia e a Tecnologia (project FCT-PTDC/ BIA-OUT/ 32077/ 2017-IC\&DT-LISBOA-01-0145-FEDER) and by the Simons Foundation (Simons Collaboration on the Global Brain \#543009).

\section*{Appendix}
\addcontentsline{toc}{section}{Appendix}
\setcounter{figure}{0}
\renewcommand\thefigure{A.\arabic{figure}}
\renewcommand{\thesubsection}{\Alph{subsection}}

\subsection{Reference table of model parameters} \label{appendix:ref-table}

% \singlespacing
\renewcommand{\arraystretch}{2}
\begin{center}
\begin{tabular}{||c | c||} 
 \hline
 \textbf{Variable(s)} & \textbf{Description} ($X,Y\in\{E, I\}$ indicate population identity) \\ [0.5ex] 
 \hline\hline
 $V_i(t), V_i^X(t)$ & Voltage of neuron $i$ (single-pop. or for pop. $X$) \\ 
 \hline
 $T_i, T_i^X$ & Threshold of neuron $i$ (single-pop. or for pop. $X$) \\ 
 \hline
 $\tau$ & Membrane time constant (and simulation timescale) \\
 \hline
 $\tau_s$ & Synaptic time constant \\
 \hline
 $\mu$ & Single-neuron cost (``soft'' refractory period) \\
 \hline
 $c(t)$ & Network input \\
 \hline
 $x(t)$ & Filtered network input \\
 \hline
 $s(t), s^X(t)$ & Spike trains (single pop. or for pop. $X$)\\
 \hline
 $r(t), r^X(t)$ & Filtered spike trains (single pop. or for pop. $X$ \\
 \hline
 $E_i, E^{XY}_i$ & Encoder weights (single-pop. or from latent $Y\rightarrow X$) \\
 \hline 
 $D_j, D_j^X$ & Decoder weights (single-pop. or for latent $X$) \\
 \hline
 $\mathcal{D}^X(x)$ & Decoder function (single-pop. or for latent $X$) \\
 \hline
 $W_{ij}, W^{XY}_{ij}$ & Recurrent weight from $j\rightarrow i$ (single-pop. or for $Y\rightarrow X$) \\
 \hline
 $y(t), y^X(t)$ & Latent variable output (single-pop. or for population $X$) \\
 \hline
 $F_i, F_i^X$ & Feedforward (input) weight (single-pop. or for latent $X$) \\
 \hline
 $B(x) = f_{cvx}(x) + y$ & Population boundary function \\
 \hline
 $I()$ & Indicator function ($I(z)=\infty$ if $z\geq0$, $I(z)=0$ otherwise) \\
 \hline
 $\sigma_\beta()$ & Sigmoid function with slope parameter $\beta$ \\
 \hline
\end{tabular}
\end{center}
% \doublespacing

\subsection{Networks with randomly-distributed parameters}\label{appendix:random}

We emphasized in the main text that, while all presented example populations were {\it tuned} to have a particular population-level boundary, our framework is fully general. This means that any spiking network with low-rank connectivity and sign constraints (reflecting inhibitory or excitatory populations) will have a well-defined boundary. The mathematical reason is that each neuron's voltage inequality defines a half-space of subthreshold voltages, and the intersection of all of these half-spaces always generates a convex subthreshold set, which is bordered by a convex boundary.

\begin{figure}[t!]
\centering
  \includegraphics[scale=1]{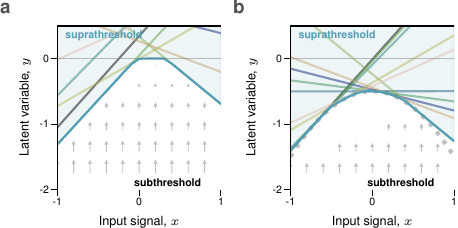}
  \caption{\label{Random-Ibounds}
    A rank-1 inhibitory population with randomly-distributed parameters.
    \textbf{a, b}: The thresholds of multiple ($N=8$, colored lines) neurons are visualized in input-output space, delineating stable, negative-convex boundaries. \textbf{a}: For each neuron $i$, parameter values were drawn as independent random variables from Gaussian distributions of the form $F_i\sim\mathcal{N}(0, 1)$, $E_i\sim\mathcal{N}(1, \tfrac{1}{4})$, and $T_i\sim\mathcal{N}(\tfrac{1}{4}, \tfrac{1}{16})$. Two properties are apparent from this ``random'' network --- (i) silent neurons: the population boundary is composed of only $3$ neurons, with the other $5$ remaining silent; (ii) silent activity regimes: for $x=0$ to $x=0.25$, the population boundary sits at $y=0$, meaning that the network will be fully silent. \textbf{b}: Same network from panel (\textbf{b}), but now with each threshold parameter $T_i$ tuned such that each neuron's boundary is tangent to the concave boundary (gray dotted line) as in \Fig{Ibounds1}. Tuning the thresholds largely solves the problems from panel \textbf{a} with silent neurons and activity regimes. However, neurons are not equally-distributed along the boundary in this case, and so each input level will have varying amounts of coding redundancy.
  }
  \end{figure}

Here, we demonstrate what happens when the neural parameters are not tuned to form a particular boundary. In \Fig{Random-Ibounds}a, we show a network of $N=8$ inhibitory neurons, where each neurons has randomly chosen values for input weights, $F_i$, latent variable encoding weights, $E_i$, and threshold $T_i$ (note that we are only interested in the shape of the boundary here, so the values of $D_i$ are not relevant). Two things are apparent. First, there are fully silent neurons, and actually the population boundary is only composed of $3$ of the $8$ neurons in this particular input domain. Second, there are values of $x$ for which the network itself is silent ($0<x<0.25$).% \footnote{\ckm{I adjusted the numbers to match the figure!}}

To avoid these degenerate regimes, we can simply tune each neuron's threshold $T_i$ (equivalent to tuning a background input current to each neuron) such that it is tangent to a desired boundary (\Fig{Random-Ibounds}b, gray dotted line). As shown in \Fig{Random-Ibounds}b, this effectively removes the degeneracy of the fully random network, and ensures that all neurons participate in forming the boundary. However, we now have a case in which different areas of the boundary may have different amounts of redundancy, and some areas with mismatches to the desired boundary. We thus conclude that (smaller) networks with randomly-distributed parameters, though still able to form population boundaries, are not ideal for approximating arbitrary boundaries. Though we used the inhibitory population as an example, these results are applicable to the excitatory population and its respective population boundary as well.

\subsection{Network parameters in the single rank-1 population} \label{appendix:Inh-params}

For the single rank-1 population with 1-d input and 1-d output, neural parameters were chosen such that each neuron was tangent to a particular boundary curve, denoted $y = -f_{cvx}(x)$. Given a network of $N$ neurons, boundaries were distributed uniformly along the curve in a particular input interval $[x_A, x_B]$, resulting in a tangent point $(x_i, y_i)$ associated with each neuron. Then, each neuron's voltage equation (\Eq{IV}) at $V_i=T_i$ was used to compute the parameters $F_i$, $E_i$, and $T_i$. Given the redundancy of the three parameters, we arbitrarily set the encoding weight, $E_i$, to be $1$ for each neuron. We thus have
\begin{align}
    E_i &= 1,\\
    F_i &= -\frac{d}{dx}f_{cvx}(x_i), \\
    T_i &= F_ix_i + E_iy_i.
\end{align}
Then, the decoding weight $D_i$ was chosen to achieve a particular jump size in the boundary at each value, and with sign constraint according to the population identity ($D_i<0$ for inhibition and $D_i>0$ for excitation).

To give a concrete example, let's consider the boundary in \Fig{Ibounds1}b,c, with boundary function $y = -x^2 - \tfrac{1}{2}$. One neuron, say neuron $i=3$, is tangent to the curve at the point ($x_i=-0.5, y_i = -0.75$). This neuron's parameters were set to
\begin{align}
    E_i &= 1,\\
    F_i &= -\frac{d}{dx} f_{cvx}(x_i) = 2x_i = -1, \\
    T_i &= F_ix_i + E_iy_i = -0.25,\\
    D_i &= -0.35.
\end{align}
Note the negative threshold, with the interpretation that the neuron is spontaneously active due to an additional background current (see \Eq{IODE}). To constrain voltages to all be on the same scale, in practice we use a default threshold $T^0_i=1$ for all neurons, and include a bias current $b_i = T^0_i - T_i$. For the case of higher, $M$-dimensional input, as in \Fig{Ibounds1}e and \Eq{vboundND}, the procedure is analogous: a tangent point is designated for each neuron in $M+1$ dimensional space, and the feedforward weights simply become a $M$-dimensional vector with the gradient of the function $-f_{cvx}(\mathbf{x}_i)$.

\subsection{Network parameters in the rank-2 EI network} \label{appendix:EI-params}

Following a similar methodology as above for the single rank-1 population, here we consider fitting neural parameters of each population to a boundary in $(x, y^E, y^I)$-space. Specifically, we denote the inhibitory boundary as $y^I = -f^I_{cvx}(x, y^E)$ and the excitatory boundary as $y^E = -f^E_{cvx}(x, y^I)$. Taking the inhibitory boundary as an example, neurons' tangent points were distributed in a 2d grid in intervals $[x_A,x_B]$ and $[y^E_A,y^E_B]$, such that each neuron had an associated point $(x_i, y^E_i, y^I_i)$. Then, as before, encoders for self-connections, $E^{II}$, were set arbitrarily to $1$ for all neurons. All parameters thus were set as
\begin{align}
    E^{II}_i &= 1,\\
    F^I_i &= -\frac{\partial}{\partial x} f_{cvx}(x_i, y^E_i), \\
    E^{IE}_i &= -\frac{\partial}{\partial y^E} f_{cvx}(x_i, y^E_i), \\
    T^I_i &= F^I_ix_i + E^{IE}y^E_i + E^{II}_iy^I_i.
\end{align}
An analogous procedure was followed for the excitatory population. Once again, in principle decoders could be set arbitrarily following sign conventions ($D^I_i<0$, $D^E_i>0$). However, we found that in practice, decoders could be optimized to ensure the two populations take turns in spiking. We generally followed a heuristic that the effective direction that the decoders point towards should aim to counteract the leak dynamics at the boundary crossing point --- in other words, $D^I/D^E \approx y_*^I/y_*^E$, where $(y_*^E,y_*^I)$ is the crossing point of the two boundaries.

\subsection{Difference of convex decomposition} \label{appendix:diff-cvx}

The expressibility of difference of convex functions has been explored in various works, perhaps starting with \citep{hartman1959functions}, and has a rich history of applications in optimization \citep{horst1999dc, yuille2003concave, bavcak2011difference, lipp2016variations}. Due to its importance to the results presented here, we provide a brief overview of this work.

To begin, we review Theorem 1 given in \citep{yuille2003concave} which proves that any function with a bounded second derivative can be written as the difference of two convex functions (or equivalently as the sum of a convex and concave function), and also provides a recipe for this decomposition:

\vspace{0.2cm}

\noindent\textbf{Theorem} \citep{yuille2003concave}. Let $f(x)$ be an arbitrary continuous function with a bounded second derivative $d^2f(x) / dx^2$. Then we can always decompose it into the difference of two convex functions.

\vspace{0.2cm}

\noindent\textbf{Proof}. We begin by selecting an arbitrary convex function $g(x)$, whose second derivative is bounded below by a positive constant $\epsilon>0$. Then there exists a positive constant $\lambda$ such that the second derivative of $q(x)=f(x) + \lambda g(x)$ is non-negative. Consequently, $q(x)$ is convex. We then define a second convex functions, $p(x) = \lambda g(x)$, and confirm that $f(x) = q(x) - p(x) = f(x) + \lambda g(x) - \lambda g(x)$.

\vspace{0.2cm}

\begin{figure}[t!]
\centering
  \includegraphics[scale=1]{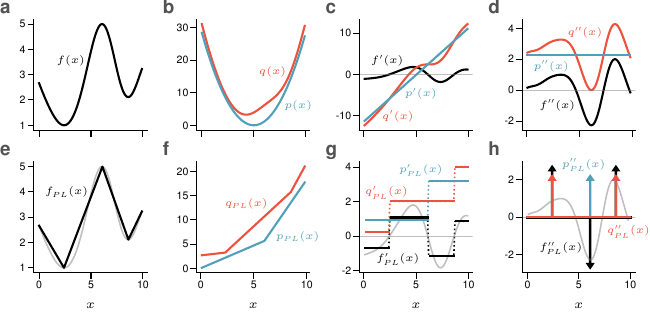}
  \caption{\label{diff-convex}
    Difference of convex function decomposition. \textbf{a-d}: A smooth, $C^2$ function $f(x)$ (\textbf{a}) is decomposed into the difference of two convex functions $q(x)$ and $p(x)$ (\textbf{b}); first- (\textbf{c}) and second-order (\textbf{d}) gradients show that $q'(x)$ and $p'(x)$ are monotone, and $q''(x)$ and $p''(x)$ are non-negative, confirming them as convex.
    \textbf{e-h}: A continuous, piecewise-linear function $f_{PL}(x)$ (\textbf{e}, black curve) is given as an approximation to the smooth curve in (\textbf{a}; gray curve in \textbf{e}), and is decomposed into the difference of two piecewise-linear convex functions $q_{PL}(x)$ and $p_{PL}(x)$ (\textbf{f}); first- (\textbf{g}) and second-order (\textbf{h}) gradients show that $q_{PL}'(x)$ and $p_{PL}'(x)$ are piecewise-constant and monotone, and $q_{PL}''(x)$ and $p_{PL}''(x)$ are non-negative, and consist of a series of scaled delta functions.
  }
  \end{figure}

A demonstration of this decomposition is shown in \Fig{diff-convex}a-d, in which a smooth, non-convex function $f(x)$ (\Fig{diff-convex}a) is decomposed into the difference of $q(x) = f(x) + \lambda x^2$ (\Fig{diff-convex}b, red) and $p(x) = \lambda x^2$ (\Fig{diff-convex}b, blue). As illustrated in \Fig{diff-convex}c,d, the derivatives of these two convex functions are monotonic, and their second derivatives are strictly non-negative, thus confirming them to be convex.

We observe that this theorem and proof apply perfectly well when $f(x)$ is continuous and piecewise-linear. However, the theorem leaves open how to choose $g(x)$ and how to work with the piecewise-linear functions required in the spiking framework. Fortunately, piecewise-linear functions can be uniquely decomposed into two piecewise-linear convex functions  \citep{melzer1986expressibility, kripfganz1987piecewise, siahkamari2020piecewise}. To illustrate the main idea, we present an additional proof.

\vspace{0.2cm}

\noindent\textbf{Theorem}. Let $f(x)$ be an arbitrary \emph{piecewise-linear} continuous function. Then we can always decompose it into the difference of two piecewise-linear convex functions.

\vspace{0.2cm}

\noindent\textbf{Proof}. We consider $f(x)$ in a particular interval of $x\in [a,b]$. Since $f(x)$ is piecewise linear, its subdifferential is piecewise constant, with a set of $n$ ``edge points'' $x_1$ to $x_n$ (including the boundary points) which denote the discontinuities, and a set of ``jumps'' $v_1$ to $v_{n-1}$, where $v_i$ indicates how much $f'(x)$ jumps at $x_i$, i.e. $v_i = (f(x_{i+1}) - f(x_i)) / (x_{i+1} - x_{i})$. We note that these sets of points and jumps will fully describe the function $f(x)$, such that we can write it as
\begin{align}
    f(x) = \sum_{i=1}^n H(x - x_i) v_ix,
\end{align}
where $H()$ denotes the Heaviside function. We now wish to decompose $f(x)$ into two piecewise-linear convex functions $q(x)$ and $p(x)$ using the same set of edge points. This is straightforward to do if we group the edge points based on the sign of each jump, $v_i$. We can then write the two functions as
\begin{align}
    q(x) = \sum_{i=1}^n H(x - x_i) H(v_i) v_ix\\
    p(x) = \sum_{i=1}^n H(x - x_i) H(-v_i) |v_i|x.
\end{align}
It is then straightforward to show that $f(x) = q(x) - p(x)$.

\vspace{0.2cm}

This second approach is demonstrated in \Fig{diff-convex}e-h, in which a piecewise-linear, non-convex function $f(x)$ (\Fig{diff-convex}e) is decomposed into the difference of two piecewise-linear, convex functions $q(x)$ (\Fig{diff-convex}f, red) and $p(x)$ (\Fig{diff-convex}f, blue). As illustrated in \Fig{diff-convex}g, the piecewise-constant gradient of the original function is decomposed into two piecewise-constant, monotonic functions which account for the positive and negative jumps in the gradient. The second-order subdifferential of $f(x)$ consists of a series of delta functions at the edge points. We now see that the two convex functions contain the same edge points, but they are strictly positive.

The above approach is in fact exactly what was used for the results in \Fig{EI-function-approx}. Once the boundary functions are determined as a function of $x$, the last step is to define the full boundary surfaces in $(x, y^E, y^I)$-space. For simplicity, and as already mentioned in the main text, we assume that the $I$ boundary is linear in $y^E$, and that the $E$ boundary is linear in $y^I$, such that the two boundaries can be decomposed into the functional forms
\begin{align}
    y^I = -p(x) - 2y^E,\\
    y^E = -q(x) - y^I.
\end{align}
Then, solving for $y^E$, we confirm that
\begin{equation}
y^E = q(x) - p(x).
\end{equation}

\subsection{Noise and irregular firing in the inhibitory network}\label{appendix:noise}

\begin{figure}[t!]
\centering
  \includegraphics[scale=1]{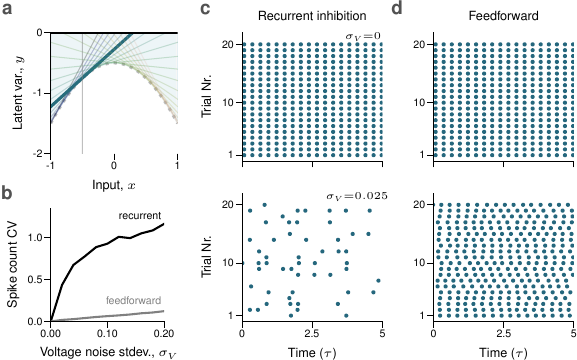}
  \caption{\label{Noise-Irregularity}
    Noise and irregular firing in the inhibitory population.
    \textbf{a}: The thresholds of multiple neurons (colored lines, shown for $N=20$) neurons are visualized in input-output space, delineating stable, negative-convex boundaries. The spike trains of one neuron (blue line) were analyzed for a fixed input $x=-0.5$ (gray vertical line). \textbf{b}: Spike count coefficient of variation (CV) plotted as a function of the voltage noise standard deviation for the default inhibitory network with recurrent inhibition (black) and comparison with a feedforward variant of the network (identical except that recurrent connections were set to zero). Spike count CV was computed over $500$ trials of a constant stimulus $x=-0.5$ for $5\tau$ time steps.
  }
  \end{figure}

In Section \ref{noisecontrol}, we illustrated how small amounts of input current noise lead to irregular firing, but it was left unspecified how much noise is actually needed. To quantify the relationship between noise and irregularity, we simulated the inhibitory population from \Fig{simulnoise} for varying noise levels, and compared it with an equivalent feedforward population in which recurrent weights were set to zero. Focusing on one particular neuron (\Fig{Noise-Irregularity}a, blue line) and input level (gray vertical line), we simulated many trials of the network and recorded spike trains. Then, computing the spike count coefficient of variation (CV) across trials, we show that the recurrent network architecture presented here results in a substantial amount of irregularity even for small amounts of noise (\Fig{Noise-Irregularity}b). The fact that the variability in the network far exceeds the feedforward network demonstrates that the irregularity is not simply due to input noise, but rather reflects the combination of the small amount of noise with the recurrent architecture. Example rasters over trials are shown for two particular noise levels for the two networks (\Fig{Noise-Irregularity}c,d).

\subsection{Synaptic dynamics} \label{appendix:synaptic-dynamics}

In Sections 2 and 3, we impose two very strict requirements on the synaptic dynamics: (i) we assume synaptic currents to be instantaneous, and (ii) we assume that inhibitory neurons always fire before excitatory neurons. The notion of instantaneous communication in a spiking network simulation with discrete time steps is tricky --- technically speaking, even in models without synaptic delays, a delay equivalent to the simulation time step, $dt$, is implicit. Event-based simulators circumvent this problem by only integrating voltages up to the next spike. To keep the simplicity of a discrete-time simulation, we instead impose that only one neuron spikes per time step. For the single inhibitory population, the rationale is straightfoward --- when the first inhibitory neuron fires a spike, it will inhibit all other neurons in the network, and thereby prevent other neurons from spiking (provided $dt$ is sufficiently small). For the EI network, an excitatory spike may instead drive other excitatory and inhibitory neurons above threshold. Here, we make the important assumption that inhibitory neurons will generally be driven above threshold faster than other excitatory neurons, and thereby are able to prevent additional excitatory spikes (see Discussion).

In Section 4, we eliminate these ad hoc rules and instead introduce finite synaptic dynamics, denoted with the variable $s_i(t)$, in the form of exponential decay or square pulses. With this modification, the voltage equation becomes
\begin{equation}
\dot{V}_i(t) = -V_i(t) + F_i c(t) + \sum_{j=1}^N W_{ij} s_j(t),\label{IODE_slow}
\end{equation}
where $s_j(t)$ is the series of postsynaptic currents due to neuron $j$.

However, we generally keep the negative diagonal terms of the connectivity instantaneously fast, as they correspond to voltage resets. This corresponds to synaptic input of the form
\begin{equation}
\sum_{j=1}^N W_{ij}s_j(t) - \mu s_i(t).
\end{equation}

\subsection{Code availability} \label{appendix:code-avail}

Code used to generate all figures of this paper is available on GitHub at \url{https://github.com/wpodlaski/funcapprox-with-lowrank-EI-spikes}. This repository also contains tutorial notebooks for visualizing latent boundaries and function approximation.

%%%%%%%%%%%%%%%%%%%%%%%%%%%%%%%%%%%%%%%%%%%%%%%%%%%%%
%%%%%%%%%%%%%%%%%%%%%%%%%%%%%%%%%%%%%%%%%%%%%%%%%%%%%
\newpage
\bibliographystyle{apalike}
\addcontentsline{toc}{section}{References}
\bibliography{refs} 

\end{document}